\documentclass[acmsmall,screen]{acmart}

\usepackage{multirow}
\usepackage{graphicx}
\usepackage[table,xcdraw]{xcolor}
\usepackage{color}
\usepackage{tabularray}
\usepackage{tcolorbox}
\usepackage[utf8]{inputenc}
\usepackage{censor}
\usepackage{caption}  
\usepackage{booktabs} 
\usepackage{soul}
\usepackage{subfigure}
\usepackage{tcolorbox}
\tcbuselibrary{listings, skins, breakable}
\usepackage{listings}
\usepackage{xcolor}
\usepackage{enumitem}
\usepackage{tabularx}

\AtBeginDocument{%
  }

\setcopyright{acmlicensed}
\copyrightyear{2026}
\acmYear{2026}
\acmDOI{XXXXXXX.XXXXXXX}

\newtcolorbox{llmprompt}[1][]{
    colback=gray!5!white,
    colframe=gray!60!black,
    fonttitle=\bfseries,
    title=Prompt provided to LLM,
    breakable,
    sharp corners,
    boxrule=0.4pt,
    enhanced,
    #1
}

\begin{document}

\title[Revisiting Vulnerability Patch Identification on Data in the Wild]{Revisiting Vulnerability Patch Identification \\
on Data in the Wild}

\author{Ivana Clairine Irsan}
\email{ivanairsan@smu.edu.sg}
\affiliation{
  \institution{Singapore Management University}
  \country{Singapore}
}

\author{Ratnadira Widyasari}
\email{ratnadiraw@smu.edu.sg}
\affiliation{%
  \institution{Singapore Management University}
  \country{Singapore}
}

\author{Ting Zhang}
\authornote{Ting Zhang is the corresponding author.}
\email{ting.zhang@monash.edu}
\affiliation{%
  \institution{Monash University}
  \country{Australia}
}

\author{Huihui Huang}
\affiliation{%
  \institution{Singapore Management University}
  \country{Singapore}
}
\email{hhhuang@smu.edu.sg}

\author{Ferdian Thung}
\affiliation{
  \institution{Singapore Management University}
  \country{Singapore}
}
\email{ferdiant.2013@smu.edu.sg}

\author{Yikun Li}
\affiliation{
  \institution{Singapore Management University}
  \country{Singapore}
}
\email{yikunli@smu.edu.sg}

\author{Lwin Khin Shar}
\affiliation{
\institution{Singapore Management University}
\country{Singapore}
}
\email{lkshar@smu.edu.sg}

\author{Eng Lieh Ouh}
\affiliation{
\institution{Singapore Management University}
\country{Singapore}
}
\email{elouh@smu.edu.sg}

\author{Hong Jin Kang}
\affiliation{%
  \institution{University of Sydney}
  \country{Australia}}
\email{hongjin.kang@sydney.edu.au}

\author{David Lo}
\affiliation{%
  \institution{Singapore Management University}
  \country{Singapore}
}
\email{davidlo@smu.edu.sg}

\renewcommand{\shortauthors}{Irsan et al.}

\begin{abstract}

Attacks can exploit zero-day or one-day vulnerabilities that are not publicly disclosed. To detect these vulnerabilities, security researchers monitor development activities in open-source repositories to identify unreported security patches.
The sheer volume of commits makes this task infeasible to accomplish manually. Consequently, security patch detectors commonly trained and evaluated on security patches linked from vulnerability reports in the National Vulnerability Database (NVD).
In this study, we assess the effectiveness of these detectors when applied in-the-wild.
Our results show that models trained on NVD-derived data show substantially decreased performance, with decreases in F1-score of up to 90\% when tested on in-the-wild security patches, rendering them impractical for real-world use. An analysis comparing security patches identified in-the-wild and commits linked from NVD reveals that they can be easily distinguished from each other. 
Security patches associated with NVD have different distribution of commit messages, vulnerability types, and composition of changes.
These differences suggest that NVD may be unsuitable as the \textit{sole} source of data for training models to detect security patches. 
We find that constructing a dataset that combines security patches from NVD data with a small subset of manually identified security patches can improve model robustness.

\end{abstract}



\received[revised]{4 February 2026}
\received[accepted]{11 March 2026}

\maketitle

\section{Introduction}
Apart from preventing vulnerabilities in their own code, developers must also mitigate threats from vulnerabilities in libraries they depend on.
To do so, developers require accurate and up-to-date information from vulnerability databases, such as the National Vulnerability Database (NVD). They may also rely on alternative vulnerability databases maintained by third-party vendors, such as Snyk\footnote{https://security.snyk.io}, Veracode\footnote{https://sca.analysiscenter.veracode.com/vulnerability-database/search}.
However, not all vulnerabilities are publicly disclosed, as library developers can patch vulnerabilities without disclosing them~\cite{prana2021out, wang2020empirical,cheng2025fixseeker}. 
Moreover, once the patch becomes public, attackers can analyze it to infer the underlying vulnerability and develop exploits~\cite{wang2019detecting,kang2022test,iannone2021toward,brumley2008automatic}. This is why detecting and reacting to such silent security patches quickly is important.

To identify undisclosed vulnerabilities, companies that maintain vulnerability databases employ security researchers who monitor open-source repositories with the aid of machine learning models~\cite{zhou2017automated, zhou2023colefunda}.
By monitoring repositories, they can detect unreported or yet-to-be-reported security patches that indicate that a vulnerability was fixed and alert developers about the vulnerabilities~\cite{sawadogo2022sspcatcher}.
However, the vast number of security patches generated across numerous repositories renders manual inspection impractical.
Machine learning-based techniques have been developed to identify security patches, where having both high precision and recall is paramount for practical usage.

These models are typically trained on commit data associated with previously identified vulnerabilities, such as those cataloged in the NVD, to predict whether an input commit has security implications.
Evaluations of benchmarks crafted from existing NVD data show that these models exhibit strong performance.

While not all vulnerabilities are publicly disclosed or reported to NVD~\cite{prana2021out}, the CVEs (Common Vulnerabilities and Exposures) cataloged on NVD are still widely used as the only source of data in constructing a dataset of security patches~\cite{zhou2023colefunda, han2024learning, zhou2021finding}. 
In this study, we analyze whether such reliance introduces a bias toward publicly disclosed vulnerabilities. For instance, we investigate whether certain types of vulnerabilities are more likely to be disclosed than others.
Recent studies~\cite{akhoundali2024morefixes, dunlap2024vfcfinder} have sought to identify security patches that are not explicitly linked or listed in the NVD. 
However, their experiments still rely on NVD entries as the starting point, meaning that their analysis ultimately remains constrained to vulnerabilities already reported in the NVD.

In the literature, there have not been prior studies that analyze the implications of using only data from reported vulnerabilities or assess their ability to detect vulnerabilities that are not publicly reported. 
In this study, we assess models trained only on NVD data and evaluate their effectiveness under a realistic evaluation setting by applying the models against a set of security patches in the wild.

We define in-the-wild security patches as patches that exist in the open source projects but are not reported to the security databases, such as NVD.
To represent these in-the-wild cases, especially from Java, we first utilize manually identified and curated patches from JavaVFC~\cite{bui2024javavfc}.
The JavaVFC dataset is constructed by monitoring the code commits made in several open-source repositories over a period of 3 years, and then manually analyzed by multiple human analysts to identify commits with security implications.
This setup is similar to the practical deployment setting of the security patch detectors in monitoring code commits, and hence, this dataset provides an opportunity for an in-the-wild analysis.
Additionally, we utilized PatchDB~\cite{wang2021patchdb} and Devign~\cite{zhou2019devign} datasets to represent in-the-wild C/C++ security patches. These datasets consist of real-world security patches collected from GitHub commits and subsequently verified by experts.

Surprisingly, the NVD-trained models exhibited significant performance degradation when applied to all in-the-wild datasets. Specifically, when evaluated on the JavaVFC dataset, we observed a substantial decrease in F1-scores of up to 90\%.
To explain performance degradation in realistic settings, we show that security patches linked to NVD entries and those identified in JavaVFC exhibit systematically different characteristics, making them easily distinguishable.

We investigate how the characteristics of the security patches differ according to their provenance.
We identify differences in the regularity of language in which the commit messages are written, as well as the composition of the commits. 
Moreover, the types of vulnerabilities differ between those found in NVD commits and those identified as in-the-wild commits.
These differences suggest that developers rarely fix some types of vulnerability in their regular development work, although these classes of vulnerabilities are heavily represented in NVD.

To enable the models to have effective performance on both commits corresponding to publicly disclosed vulnerabilities and those fixed without public disclosure, we combined the datasets and found that this enables stable performance on security patches regardless of their provenance. 
Moreover, including a small proportion of commits from the JavaVFC data is already sufficient to significantly improve the performance of the model.

Additionally, we investigate whether training models solely on code changes can enhance the robustness to the distributional shift from NVD data to Java OSS data.
Based on experiments, we observed that leveraging only code changes resulted in inferior performance compared to utilizing both commit messages and code changes.
Therefore, further studies should consider the use of both commit messages and code changes to identify silent security patches.

We also discuss the lessons learned and implications of our experiments regarding the evaluation of future security patch detectors. 

In summary, this paper makes the following contributions:
\begin{itemize}[leftmargin=1em]
    \item  \textbf{Evaluation.} We replicated state-of-the-art security patch detectors from prior work and evaluated them alongside several LLMs to assess their performance in a realistic, in-the-wild setting.
    \item \textbf{Analysis.} We performed an in-depth analysis of the differences between security patches linked from publicly disclosed vulnerabilities in the NVD entries and patches that were not disclosed. 
    \item \textbf{Mitigation.} We showed that by constructing a new training dataset based on NVD-associated security patches and a small number of unreported security patches, we can enhance the performance of a security patch detector.
\end{itemize}

The remaining sections of this paper are organized as follows. 
Section~\ref{sec:background} introduces the background of the work, including the security patch detectors and the datasets studied in our experiments.  
Section~\ref{sec:setup} presents the experimental setup. 
Section~\ref{result} analyzes the results of the experiments. 
Section~\ref{sec:discussion} discusses their implications, lessons learned, and threats to validity. 
Section~\ref{sec:related} introduces other related work. 
Section~\ref{sec:conclusion} concludes the paper and discusses future work.

\section{Background}
\label{sec:background}

\subsection{Security Patch Detection Approaches}
Prior approaches for automatically detecting security patches formulate the task as a binary classification problem. 
Given a commit consisting of its commit message and the code change, the security patch detector predicts whether the commit fixes a vulnerability or not. 
The state-of-the-art approaches~\cite{zhou2023colefunda, han2024learning, sun2023silent} train deep learning models.

In the approach proposed by Sun et al.~\cite{sun2023silent}, security patch detection is integrated into an explainable AI pipeline for silent dependency alert prediction.
Their experiments evaluated a wide range of models.
Apart from finding that the use of both the commit  message and all code changes in the commit led to the best results, 
their experimental results indicate that
CodeBERT outperforms other pretrained models in detecting security patches.

ColeFunda~\cite{zhou2023colefunda} is a framework for security patch detection, designed to identify silent fixes and provide detailed vulnerability explanations.
To detect silent fixes, ColeFunda does not utilize information from the commit message, performing classification using only the code change. 
Under a coordinated disclosure policy, developers should not reveal information about the vulnerability. 
Thus, developers would not reveal vulnerability information in the commit message.
Zhou et al. prevent their model from using commit messages to ensure effectiveness even in the absence of useful commit messages.
Instead, ColeFunda employs data augmentation before learning representations of code changes using contrastive learning. 

GRAPE~\cite{han2024learning} is a graph-based patch representation framework designed to unify vulnerability fix patch representation and enhance the understanding of patch intent and impact by extracting structural information from code.
At its core, GRAPE employs a novel joint graph structure, MCPG, to represent both the syntactic and semantic information of silent fix patches, embedding attributes for nodes and edges. A graph convolutional neural network (NE-GCN) then leverages these structural features.
Similar to ColeFunda, GRAPE does not use information from the commit message, relying solely on the code change. 
In their evaluation of GRAPE, Han et al. constructed a dataset of 2,251 commits. 
On this dataset, their experimental results show that GRAPE outperforms the baseline methods by significantly reducing false positives and omissions in security fix identification.

All of these methods, including the techniques developed for detecting silent fixes, rely on training data derived from commits associated with NVD entries. 
Moreover, their evaluation is limited exclusively to commit data from the NVD. The NVD, maintained by the U.S. National Institute of Standards and Technology, is a comprehensive repository of publicly disclosed software vulnerabilities.
While such databases contain a large quantity of data, they do not have records of vulnerabilities that were not publicly disclosed~\cite{votipka2018hackers,gorbenko2017experience}.

\subsection{Security Patch Benchmark Based on NVD Entries}
Previously proposed security patch detectors train deep learning models using security patches associated with NVD. 
In our experiments, we obtained and utilized the dataset from the ColeFunda evaluation. 
The data split used in the ColeFunda study is publicly available.\footnote{https://figshare.com/s/840e1fb94bd972829c80}
This dataset is specifically designed for the Java programming language and is derived from NVD data.

In addition to the datasets constructed for training security patch detectors in prior work, Akhoundali et al.~\cite{akhoundali2024morefixes} introduced the MoreFixes vulnerability database, an extension of the  CVEFixes dataset~\cite{bhandari2021cvefixes}. 
MoreFixes employs several heuristic methods to integrate state-of-the-art approaches for gathering CVE fix commits.
It is the largest real-world dataset of CVE vulnerabilities and their corresponding fix commits. 
It comprises 26,617 unique CVEs, categorized by Common Weakness Enumeration (CWE), from 6,945 GitHub projects across multiple programming languages.
In addition to vulnerability report information, it includes comprehensive details about each CVE up to January 2024, such as how vulnerabilities were fixed, the files and methods modified in the fix commits, and other metadata about the associated repositories.
Compared to the ColeFunda dataset, MoreFixes includes all the CVEs listed in the ColeFunda dataset. 
In total, the ColeFunda dataset contains 839 CVEs, derived from a previous study~\cite{zhou2021finding}. 
Although ColeFunda's CVEs are a subset of those in MoreFixes, the associated security patches are slightly different.
The security patches associated with MoreFixes are a superset of security patches from ColeFunda, as MoreFixes combines information from multiple sources (both NVD and Github advisories\footnote{https://github.com/advisories}).

Finally, the dataset used in the GRAPE experiments~\cite{han2024learning} is derived from that constructed by Ponta et al.~\cite{ponta2019manually}. 
This dataset contains 624 publicly disclosed vulnerabilities. 
The vulnerabilities affect 205 distinct open-source Java projects used in SAP products or their internal tools.
Among those, 12\% were not reported in NVD at the time of writing in 2019, but were manually curated by the SAP KB\footnote{https://sap.github.io/project-kb/} project, maintaining its quality.
Although the dataset by Ponta et al. contains vulnerabilities that are not reported to NVD, only 29 vulnerabilities are not assigned CVE identifiers, and all of them are publicly disclosed.
On the other hand, the JavaVFC dataset~\cite{bui2024javavfc} used in this study consists of security patches filtered from millions of random commits using a keyword-based method, followed by a rigorous curation process.

We use all three datasets in our experiments and show that these datasets ultimately suffer from the same drawback of including only publicly disclosed security patches.

\subsection{Assessing the Performance of Research Tools Under Realistic Settings}

Prior studies have indicated the importance of assessing tools beyond simple benchmarks, as it has been a widely studied topic~\cite{wang2019detecting, chakraborty2024revisiting, nesti2022evaluating, bui2024apr4vul, gong2023wfdefproxy, ancha2024utilizing, jamieson2015next, mashlakov2021assessing, raza2015practical}. 
Notably, several studies evaluated prior research tools in the wild, such as those addressing vulnerability detection~\cite{chakraborty2024revisiting} and traceability tasks that link commits to issues~\cite{rath2018traceability}.
A recent study on bug detection~\cite{richter2023train} revealed that neural bug detectors trained on artificial bugs show inferior performance when tested on real-world datasets.
This finding prompts the need to incorporate more realistic, real-world data into training processes to improve model performance in practical applications.
While our work is closely related to that of Chakraborty et al.~\cite{chakraborty2024revisiting}, who revisited the performance of deep-learning-based vulnerability detectors on the Real-Vul dataset (comprising 10 popular C/C++ projects), our study tackles a different problem. Specifically, we examine what distinguishes reported vulnerabilities from those that go unreported, and whether data from reported vulnerabilities alone suffices to uncover unreported ones. Furthermore, our approach is not restricted to a fixed set of projects, which enables a broader investigation.

\subsection{Security Patches In the Wild}
\label{sec:javaVFC}
Since the practical task of security patch detection involves monitoring the development activity of a large number of repositories, the models would have to generate predictions for each commit. 
An in-the-wild analysis should account for fixes with security implications that were made without public disclosure, which would allow us to assess the impact of training models only on NVD data.

One method of assessing the models is to apply them to security patches identified by a different method, independently of the security patches associated with NVD entries. 
Prior research has used keyword matching to curate security patch datasets~\cite{zhou2017automated}; however, these data are obtained from 8,536 projects in multiple programming languages and based on issues reported on various platforms such as GitHub, JIRA, and Bugzilla.
In other words, this dataset was crafted to contain reported vulnerabilities instead of covering real silent vulnerability patches, which were never reported to any issue-tracking or vulnerability-tracking platform.
Recently,  Bui et al.~\cite{bui2024javavfc} introduced a dataset, JavaVFC, curated from commits of more than 34,000 open-source Java repositories on GitHub.
The authors provide two versions of the dataset: JavaVFC, which contains 784 manually verified security patches, and JavaVFC-extended, which comprises 16,837 automatically identified security patches.
These commits were collected over a three-year period, from February 2021 to February 2024.
Security patches were identified using a keyword-based heuristic to filter commits with security-related terms in their messages and were manually labeled with a high level of agreement between human analysis.
Our experiments use the human-curated dataset (JavaVFC) to assess the security patch detectors. 

\section{Experimental Setup}
\label{sec:setup}
In this section, we explain the experimental setup of our study.

\subsection{Overview}

\begin{figure*}[]
    \centering
    \includegraphics[width=1.0\textwidth]{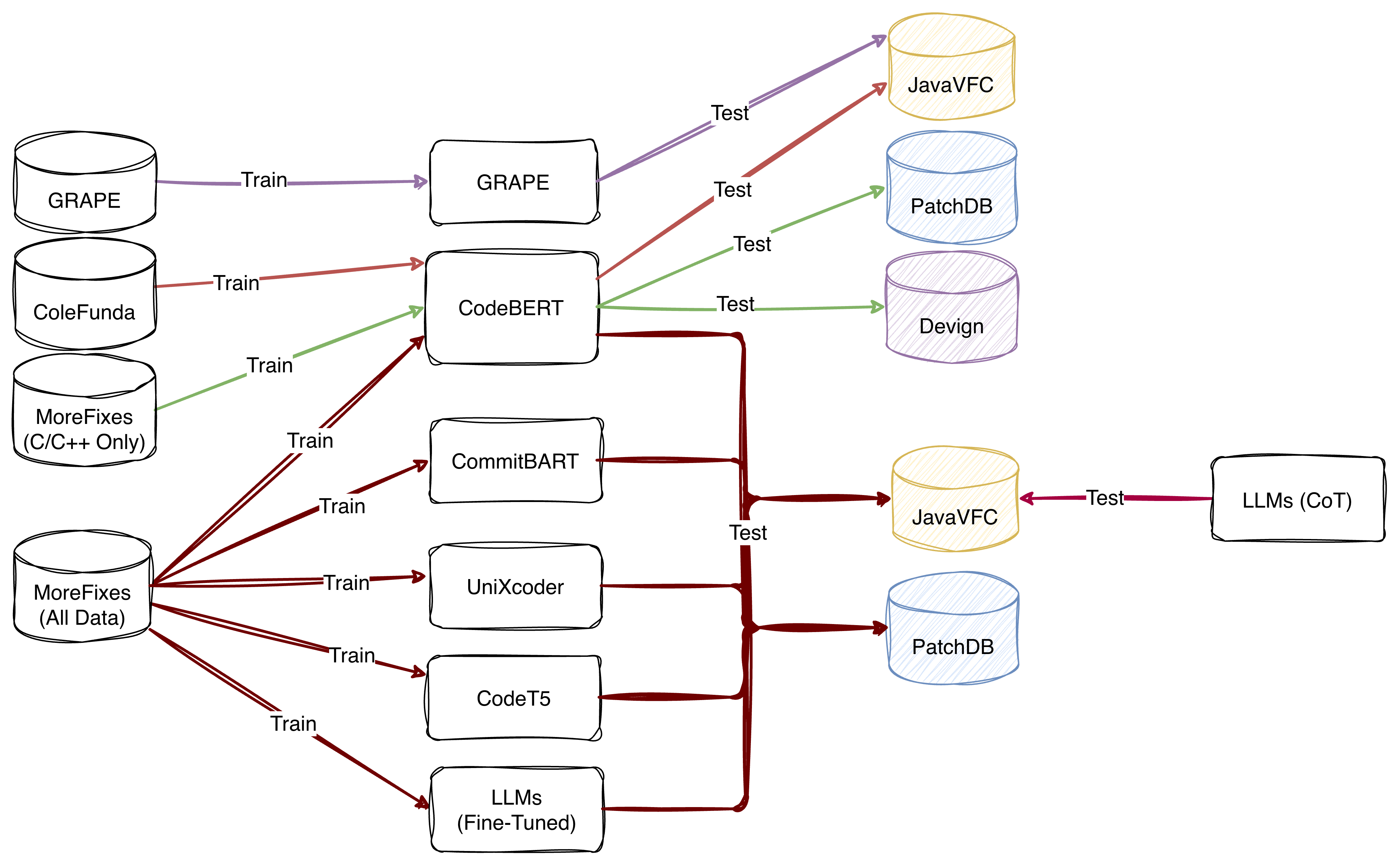}
    \caption{Overview of the realistic evaluation of the models trained on NVD-linked security patches. We assess their performance by applying them to in-the-wild patches derived from open-source repositories
    }
    \label{fig:overview}
\end{figure*}

Our experiments are driven by the following research questions (RQs):

\vspace{4px}
\noindent\textbf{RQ1. Do models trained only on publicly disclosed security patches generalize to undisclosed security fixes?} 
This research question investigates the effectiveness of tools trained on NVD data when applied to commits fixing vulnerabilities or security patches that were not publicly disclosed.
The overview of the experiment conducted in this study is presented in Figure \ref{fig:overview}.  

We begin by training models on datasets~\cite{akhoundali2024morefixes, zhou2023colefunda, han2024learning}, derived from the NVD, and then evaluate these models on the JavaVFC~\cite{bui2024javavfc},  PatchDB~\cite{wang2021patchdb}, and Devign~\cite{zhou2019devign} datasets, which contain fixes that were not publicly disclosed, to assess the models under realistic evaluation.

To investigate this question, we run our experiments on a range of models and datasets. 
Our analysis focuses on CommitBART~\cite{liu2024automated}, CodeBERT, CodeT5, and UniXcoder. 
CommitBART was specifically designed for commit-related tasks and has been previously evaluated for security patch detection. In contrast, CodeBERT achieved the highest performance in the recent experiments conducted by Sun et al.~\cite{sun2023silent}. Furthermore, we included CodeT5 and UniXcoder, as these models demonstrated competitive performance relative to CodeBERT in a study by Ding et al.~\cite{ding2024vulnerability}.

Furthermore, we included GRAPE~\cite{han2024learning}, an approach recently released in 2024, as a baseline. According to its original paper, GRAPE outperforms other commonly known approaches such as PatchRNN~\cite{wang2021patchrnn}, GraphSPD~\cite{wang2023graphspd}, and SPI~\cite{zhou2021spi} by 8-25\%. Therefore, we excluded those approaches from our comparisons, as we believe that evaluating against GRAPE provides a more effective and resource-efficient benchmark, given that it already surpasses those models significantly.

However, GRAPE's applicability is restricted because it relies on Joern\footnote{\url{https://joern.io}} for preprocessing, and Joern supports only 10 of the 81 programming languages found in the MoreFixes dataset~\cite{akhoundali2024morefixes}.
As a result, we evaluated GRAPE solely on its original dataset and the JavaVFC dataset.

Additionally, we conducted experiments with various LLMs to assess whether larger models could effectively detect security patches in the wild.
For these experiments, we evaluated a diverse set of open-source LLMs, including DeepSeek-R1~\cite{guo2025deepseek}, Qwen3~\cite{yang2025qwen3}, Llama-3.1~\cite{touvron2023llama}, DeepSeek Coder~\cite{guo2024deepseek}, as well as proprietary OpenAI models such as GPT-4.1, o3-mini, and GPT-4.1-mini.
For open source models, we prioritize those developed for reasoning, as such models have been shown to produce positive transfer effects on a wide range of downstream tasks~\cite{huan2025does, ma2025general, xu2025towards}.
Therefore, we opt to use models such as DeepSeek-R1\footnote{https://huggingface.co/deepseek-ai/DeepSeek-R1-Distill-Qwen-7B}  and Qwen3\footnote{ https://huggingface.co/Qwen/Qwen3-8B}, while for the OpenAI models we studied multiple variants from general tasks and reasoning series.
In addition, we included Llama-3.1 and DeepSeek Coder to represent models that were not specifically developed for reasoning, serving as comparative baselines.

In addition, we also investigate whether exposure to raw code during pre-training without corresponding security labels impacts the ability of an LLM to identify security patches in a zero-shot setting (i.e., without fine-tuning). We utilized DeepSeek Coder\footnote{https://huggingface.co/deepseek-ai/deepseek-coder-7b-instruct-v1.5} and Llama3.1\footnote{https://huggingface.co/meta-llama/Llama-3.1-8B-Instruct}, as these models have knowledge cutoff dates that align with the JavaVFC dataset. We divide the data into samples released before and after each model’s knowledge cutoff and compare the performance across these subsets. We then performed a statistical significance test to assess whether the observed differences are significant.

Chain-of-thought prompting was used based on recent studies~\cite{ding2024vulnerability} that assess the capabilities of LLM to detect vulnerable functions.
In that study, the application of chain-of-thought reasoning outperformed the fine-tuning of GPT-3.5 and GPT-4 for vulnerability detection.
Although vulnerability detection and security patch identification are not identical tasks, we argue that they share similar traits, requiring LLMs to analyze the semantics of code.
Therefore, we adopted the chain-of-thought technique, as it demonstrates better performance compared to the more resource-intensive fine-tuning approach.

Specifically, we use the following prompt:

\begin{llmprompt}
You are a security expert that is good at static program analysis.

Please analyze the following code:

\textasciigrave\textasciigrave\textasciigrave
[MSG] <commit message> [/MSG] [DIFF] <commit diff> [/DIFF]
\textasciigrave\textasciigrave\textasciigrave

Please indicate your analysis result with one of the options: 

(1) YES: A security vulnerability detected.

(2) NO: No security vulnerability.

Put it in this output format: <answer> [YES | NO] </answer>

Output example if a security vulnerability is detected: <answer> YES </answer>

Make sure to follow the output format explicitly (EXPLICITLY!!!) in your response.

Let's think step-by-step.
\end{llmprompt}

Furthermore, to complement our results, we fine-tuned three popular LLMs, i.e., DeepSeek Coder\footnote{https://huggingface.co/deepseek-ai/deepseek-coder-7b-base-v1.5}, Llama3.1\footnote{https://huggingface.co/meta-llama/Llama-3.1-8B}, and Qwen3\footnote{https://huggingface.co/Qwen/Qwen3-8B}, using the NVD-derived MoreFixes dataset.
We fine-tuned the selected LLMs using Parameter-Efficient Fine-Tuning (PEFT) to adapt them to the downstream task of security patch identification. Specifically, we applied Low-Rank Adaptation (LoRA), which freezes the pre-trained model weights while injecting trainable rank-decomposition matrices into the transformer layers. 
To support the binary classification objective, we replaced the language modeling head with a classification head. The models were trained for three epochs with the objective of minimizing the cross-entropy loss. To optimize training efficiency and prevent overfitting, we adopted a step-based evaluation strategy combined with an early stopping mechanism. The F1-score was used as the primary optimization metric.

Following the fine-tuning process, we evaluated their performance on the JavaVFC and PatchDB datasets to assess their generalizability to in-the-wild security patches. 
This approach allowed us to determine whether the specialized knowledge gained from NVD data effectively transfers to security patch identification in real-world, unreported scenarios.

\vspace{4px}
\noindent\textbf{RQ2. How do the unreported security patches differ from security patches linked from NVD?}
This research question aims to investigate the differences between the security patches.
We analyze the patches in terms of their commit messages, vulnerability types, and the compositions of the commits. 

To assess whether security patches are distinguishable based on their commit messages, we conducted several analyses. First, we performed a classification experiment using XGBoost to determine whether security patches originated from NVD or Java OSS, represented by the MoreFixes and JavaVFC datasets, respectively. Second, we leveraged perplexity scores to compare whether commit messages from MoreFixes and JavaVFC are more frequently represented in the data used to train a pretrained code model. Third, we contrasted the CWE classes reported in MoreFixes with those found in a sample of JavaVFC data. Additionally, we analyzed the overlap between MoreFixes and JavaVFC and conducted intra-project experiments to ensure that our results were not biased by project-specific differences. Finally, we examined the composition of commits to gain a deeper understanding of the characteristics of commits originating from MoreFixes and JavaVFC.

\vspace{4px}
\noindent\textbf{RQ3. Does combining data from both sources, NVD and Java open source data, lead to better and more robust performance? } 
This research question examines whether we can understand 
how the number of human-labeled samples influences security patch detection. 
Given the limited availability of manually curated data, we aim to determine whether it is sufficient to significantly improve the model's performance.
To address this research question, we introduce data from JavaVFC for training. Then, to assess the impact of including these additional data, we incrementally increase the amount introduced and compare the resulting performance across increments.

\subsection{Data}
\label{sec:data}
In this study, we use two types of data to assess the security patch detectors.

\noindent\textbf{NVD data.} First, our experiments use datasets from previous studies that analyzed or experimented with data from the NVD:
\begin{enumerate}[leftmargin=2em]
    \item \textbf{ColeFunda - Java Security Patches}: This dataset was utilized in prior studies~\cite{zhou2023colefunda,han2024learning}. It consists of security patches linked to NVD reports.  
    As the ColeFunda approach was designed for Java, commits from non-Java projects are filtered out.
    To obtain the negative samples, non-fix commits are collected from the same projects as the positive samples. 
    Details regarding the number of training and testing data are presented in Table~\ref{tab:dataset-statistics}.
ColeFunda adopts an imbalanced data setting in their experiment, which we followed by using the same data split provided.

    \item \textbf{GRAPE - Java Security Patches}:
     The GRAPE dataset employs a balanced setting, which we replicated by using their dataset and splitting it with an 8:1:1 ratio for training, validation, and testing.
    
    \item \textbf{MoreFixes - Security Patches}: 
The MoreFixes dataset~\cite{akhoundali2024morefixes} is constructed from security patches that can be linked to NVD entries, without filtering out projects based on a specific programming language. 
We use all available security patches from MoreFixes to create a multilingual training set. 
Unlike the ColeFunda and GRAPE datasets, MoreFixes is not limited to Java.
\end{enumerate}

\vspace{4px}
\noindent\textbf{Security Patches in the Wild.} Second, we perform an in-the-wild analysis by running experiments with data collected from open source repositories:

\begin{enumerate}[leftmargin=2em]

\item \textbf{JavaVFC}~\cite{bui2024javavfc}. As described in Section \ref{sec:javaVFC}, the patches are curated from the code commits made to Java projects over the course of 3 years, from Feb 2021 to Feb 2024. These commits were manually analyzed by multiple human security analysts to be security-relevant. 

\item \textbf{Devign - Partial}~\cite{zhou2019devign}.
The initial Devign dataset contains verified security patches from 4 popular open-source libraries: Linux, FFmpeg, Qemu, and Wireshark. However, the authors only released partial data derived from 2 projects\footnote{https://sites.google.com/view/devign}, which consists of commits from the Qemu and FFmpeg datasets. 
Commits in this dataset are written in C/C++ language.

\item \textbf{PatchDB}~\cite{wang2021patchdb}. The PatchDB dataset is a large-scale security patch dataset built by Sun Security Laboratory that contains roughly 8,000 C/C++ security patches collected from the real world.
In our study, we utilize the \textit{wild-based} dataset that consists of security patches that are collected from commits in GitHub via a nearest link search method to help find the most promising security patch candidates, which are then further verified by multiple experts.

\end{enumerate}

Each of the datasets used to represent security patches in the wild is constructed through multiple steps, including automatic filtering followed by manual checks to ensure correctness. 
JavaVFC and Devign employ keyword filtering to detect potential security patches through commit messages, while PatchDB uses a nearest link search method as the first step in identifying security patches in the wild.
These candidates are then verified by multiple experts to ensure the correctness of the labels.
Therefore, these datasets are suitable for representing unreported security patches in the wild, as they are derived from open-source projects with no link to reported vulnerabilities.

Furthermore, although these security patches are not formally reported, they undergo multiple layers of validation to ensure label quality, making them reliable representations of unreported in-the-wild security patches. Table~\ref{tab:dataset-statistics} presents detailed statistics for each dataset.


\begin{table}[t]
\centering
\caption{Information about the datasets, including the source of the data and the number of positive and negative samples.}
\label{tab:dataset-statistics}
\begin{tabularx}{\columnwidth}{lllXrr}
\toprule
\textbf{Dataset} & \textbf{Language} & \textbf{Provenance} & & \textbf{\# Sec. Patches} & \textbf{\# Non-Sec. Patches} \\
\midrule
ColeFunda        & Java              & NVD                 & & 1,436  & 839,682 \\
GRAPE            & Java              & NVD                 & & 1,068  & 1,183   \\
MoreFixes        & Multi (81 PL)     & NVD                 & & 31,883 & 0       \\
JavaVFC          & Java              & Java OSS            & & 784    & 0       \\
PatchDB          & C/C++             & C OSS               & & 7,997  & 23,742  \\
Devign (partial) & C/C++             & Qemu \& FFmpeg      & & 10,891 & 14,977  \\
\bottomrule
\end{tabularx}
\end{table}

\vspace{4px}
\noindent\textbf{Negative Samples}. 
To create training and testing datasets for security patch detection, it is essential to have both positive samples (security patches) and negative samples (non-security patches). The ColeFunda dataset already provides predefined training, validation, and test splits with both positive and negative samples. In contrast, the GRAPE dataset includes positive and negative samples, but lacks an explicit split. Therefore, we divided GRAPE’s data into an 8:1:1 ratio for training, validation, and testing.
On the other hand, both MoreFixes and JavaVFC lack negative samples. Hence, we sampled five random non-security commits for each security patch in these datasets via the GitHub API.\footnote{https://docs.github.com/en/rest} This led to five negative samples per positive sample, resulting in 3,805 non-security patch commits for JavaVFC and over 160,000 non-security patch commits for MoreFixes. Depending on the dataset used to train the security patch detection model, we employ either balanced or imbalanced settings.

Specifically, to address RQ1, we evaluate the security patch detector under both balanced and imbalanced settings. For model trained on the ColeFunda dataset, we use a 1:5 ratio of positive to negative samples in the testing data to introduce an imbalance, aligning with the context of the original study.
It is important to note that the ColeFunda dataset exhibits a much higher level of imbalance compared to our testing data. By reducing the complexity of the imbalance in the test dataset, our setting becomes more forgiving for models trained on ColeFunda data. This is because a lower imbalance ratio in the test data generally makes it easier for the model to perform well~\cite{jj2024digital, zhang2020novel, he2009learning}.
In contrast, we use a 1:1 ratio of positive to negative samples to test models trained on the GRAPE and MoreFixes datasets. This approach assesses the models' fundamental ability to identify security patches without the added complexity of data imbalance.

Overall, our settings are designed to evaluate models trained exclusively on security patches derived from reported vulnerabilities (e.g., NVD) by testing them on security patches derived from unreported vulnerabilities (i.e., in-the-wild). This creates a more realistic setting to assess whether patches from reported vulnerabilities are sufficient to detect silent security patches in-the-wild.
Additionally, we opt for a less complex positive-to-negative data ratio compared to the original datasets. This decision aligns with our primary goal: to evaluate whether the reported security vulnerabilities provide sufficient features to detect actual security patches, especially given that many security patches remain unreported. Rather than benchmarking models across various positive-to-negative ratios in reported security patches, our aim is to highlight the broader applicability of using reported patches to train effective models.

\subsection{Evaluation Metrics}
To evaluate the approaches in this study, we use precision, recall, and F1-score following previous studies~\cite{han2024learning, xu2019sentiment, zhang2020sentiment, wu2022enhancing, zhou2021spi,zhang2025revisiting}. 
These are  well-established metrics commonly used in the classification task.
We use the F-measure with $\beta$=1, also known as \texttt{F1-score} as the primary metric to compare the approaches.
The formula for precision, recall, and F1-score are presented in \ref{eqn:precision}, \ref{eqn:recall}, and \ref{eqn:f_measure}.

\begin{equation} \label{eqn:precision}
\text{Precision} = \frac{TP}{TP + FP} 
\end{equation}

\begin{equation} \label{eqn:recall}
\text{Recall} = \frac{TP}{TP + FN} 
\end{equation}

TP represents the number of true positives, FP represents false positives, and FN represents false negatives.

F$\beta$ balances precision and recall using the harmonic mean, where coefficient = 1 indicates that both precision and recall are of similar importance.

\begin{equation}
    F_\beta = \left(1 + \beta^2\right) \cdot \frac{\text{Precision} \times \text{Recall}}{\beta^2 \cdot \text{Precision} + \text{Recall}}
    \label{eqn:f_measure}
\end{equation}

Note: In this study, we set \(\beta = 1\), which is the F1-score.

\section{Results}
\label{result}

\subsection{\textit{RQ1: Do models trained only on publicly disclosed security patches generalize to undisclosed security fixes?}}\label{sec:RQ1results}

\begin{table}[ht]
\centering
\caption{Performance comparison between NVD and OSS testing datasets, showing a significant decrease in F1-score on the language-specific setting.}
\resizebox{\textwidth}{!}{
\begin{tabular}{llllcccc}
\toprule
\textbf{Model} & \textbf{Train Data} & \textbf{Language} & \textbf{Test Data (Provenance)} & \textbf{Prec.} & \textbf{Rec.} & \textbf{F1} & \textbf{$\Delta$F1} \\ \midrule
\multirow{2}{*}{CodeBERT} & \multirow{2}{*}{ColeFunda} & \multirow{2}{*}{Java} & ColeFunda (NVD) & 98.82 & 84.77 & 91.26 & -- \\
                          &                            &                       & JavaVFC (OSS) & 40.22 & 4.86 & 8.68 & \color{red}$-90\%$ \\ \midrule
\multirow{2}{*}{GRAPE}    & \multirow{2}{*}{GRAPE}     & \multirow{2}{*}{Java} & GRAPE (NVD) & 76.67 & 82.14 & 79.31 & -- \\
                          &                            &                       & JavaVFC (OSS) & 48.55 & 39.41 & 43.51 & \color{red}$-45\%$ \\ \midrule
\multirow{3}{*}{CodeBERT} & \multirow{3}{*}{MoreFixes} & \multirow{3}{*}{C/C++}  & MoreFixes (NVD) & 78.96 & 84.88 & 81.81 & -- \\
                          &                            &                         & PatchDB (OSS) & 30.61 & 86.48 & 45.22 & \color{red}$-38\%$ \\
                          &                            &                         & Devign (OSS) & 47.36 & 83.06 & 60.32 & \color{red}$-26\%$ \\ \bottomrule
\end{tabular}
}
\label{tab:result-per-language}
\end{table}

\begin{table}[t]
\centering
\caption{Performance comparison between NVD and OSS testing datasets, utilizing multiple languages as training data.}
\label{tab:result-all-language}
\begin{tabular}{lllrrrc}
\toprule
\textbf{Model} & \textbf{Train Data} & \textbf{Test Data (Provenance)} & \textbf{Prec.} & \textbf{Rec.} & \textbf{F1} & \textbf{$\Delta$F1} \\
\midrule
\multirow{3}{*}{CodeBERT}
  & \multirow{3}{*}{MoreFixes}
  & MoreFixes (NVD) & 61.60 & 83.72 & 70.97 & --                    \\
  & & JavaVFC (OSS)   & 44.39 & 44.15 & 44.27 & \color{red}{$-38\%$} \\
  & & PatchDB (OSS)   & 29.55 & 87.75 & 44.21 & \color{red}{$-38\%$} \\
\midrule
\multirow{3}{*}{CommitBART}
  & \multirow{3}{*}{MoreFixes}
  & MoreFixes (NVD) & 63.98 & 60.04 & 61.95 & --                    \\
  & & JavaVFC (OSS)   & 26.50 & 54.40 & 35.64 & \color{red}{$-42\%$} \\
  & & PatchDB (OSS)   & 32.84 & 70.48 & 44.80 & \color{red}{$-28\%$} \\
\midrule
\multirow{3}{*}{UniXcoder}
  & \multirow{3}{*}{MoreFixes}
  & MoreFixes (NVD) & 63.82 & 78.57 & 70.43 & --                    \\
  & & JavaVFC (OSS)   & 31.51 & 87.12 & 46.28 & \color{red}{$-34\%$} \\
  & & PatchDB (OSS)   & 40.86 & 76.59 & 53.29 & \color{red}{$-24\%$} \\
\midrule
\multirow{3}{*}{CodeT5}
  & \multirow{3}{*}{MoreFixes}
  & MoreFixes (NVD) & 63.15 & 71.33 & 67.00 & --                    \\
  & & JavaVFC (OSS)   & 28.30 & 88.17 & 42.85 & \color{red}{$-36\%$} \\
  & & PatchDB (OSS)   & 40.03 & 82.09 & 53.82 & \color{red}{$-20\%$} \\
\bottomrule
\end{tabular}
\end{table}

Table \ref{tab:result-per-language} and Table \ref{tab:result-all-language} summarize our main experimental results.
Table \ref{tab:result-per-language} presents the results for the models trained and tested on single-language datasets, whereas Table \ref{tab:result-all-language} details performance when the models are trained on all available data, regardless of language.

As shown in Table~\ref{tab:result-per-language}, training and testing models on specific languages yield higher F1-scores when training and testing data share the same provenance (i.e., training and testing both on NVD data). 
Using the ColeFunda dataset, as the implementation of ColeFunda is not publicly available, we trained a CodeBERT model following the approach of Sun et al.~\cite{sun2023silent} and found that we achieved a higher AUC (0.9) compared to the reported results by Zhou et al.~\cite{zhou2023colefunda} (0.8).
Next, we also reproduced GRAPE~\cite{han2024learning} by reusing the code from the publicly available replication package, achieving an F1-score of 79.31\% compared to the reported F1-score of 88.52\%.
After training the models on the NVD dataset, we evaluated them on the JavaVFC dataset. 
In our experiments, we observed a significant drop in performance in all models, with the CodeBERT model having a larger drop in performance of 90\% when compared to the GRAPE model's drop of 45\% when trained on Java language only.

In addition to Java, we also conducted an experiment with C/C++ language trained with CodeBERT to validate the generalizability of our study. To evaluate the CodeBERT model trained on NVD data, we utilized released patches from Devign and the wild-based PatchDB dataset, which is constructed from patches collected from more than 300 popular C/C++ Github repositories. 
As a result, we observed that when CodeBERT was trained and tested with NVD data in C/C++ language, the F1-score achieved 81.81\%. However, when the same model was tested on in-the-wild security patches derived from the Devign and PatchDB dataset, the performance dropped for more than 26\%.

Overall, these single-language models exhibit a more significant performance decline when evaluated on in-the-wild datasets compared to models trained on multiple languages.
These results suggest that training models on a diverse set of security patches across all available languages is more practical and beneficial for real-world applications, as it enhances the models' ability to identify in-the-wild security patches.

Next, as the prior studies~\cite{zhou2023colefunda, han2024learning} filtered out commits from non-Java projects, they used a smaller amount of training data.
To investigate whether the poor generalizability of the models stemmed from an insufficiently sized training dataset, we leveraged the more comprehensive NVD dataset, MoreFixes, in addition to the Java-only security patch datasets.
This provided a broader representation of vulnerabilities across multiple programming languages, and a much larger dataset (over 22x increase in size).
Similarly, we applied the same experimental setup to the C/C++ language to further examine the impact on security patches detection models when trained on larger NVD-based datasets and tested on in-the-wild patches.

The results in Table~\ref{tab:result-all-language} show that CommitBART~\cite{liu2024automated}, despite being more recently released, fails to outperform CodeBERT.
Specifically, CodeBERT remains superior, achieving a 14\% higher F1-score than CommitBART when trained and tested on the MoreFixes dataset.
Similar trends are observed with UniXcoder and CodeT5; both are inferior to CodeBERT on the NVD-based benchmark and exhibit a significant performance drop when tested on in-the-wild security patch datasets. 

Additionally, CodeBERT was able to achieve a high level of effectiveness on a testing data split from the MoreFixes data. 
However, while we observe that the use of a much larger dataset created from commits from diverse programming languages leads to a smaller drop in performance, the loss of effectiveness is still significant with a 38\% decrease in F1-score on the JavaVFC dataset.
The experiment on the C/C++ language also exhibits traits similar to Java, with a significant performance drop observed when CodeBERT was trained on the MoreFixes dataset and tested on a dataset representing in-the-wild C/C++ security patches. This drop was especially pronounced when tested on the larger in-the-wild dataset, PatchDB, which corresponded to a decrease of 38\%.
Therefore, we can eliminate the possibility that the poor generalizability was caused by the smaller training dataset. 
The consistent drop in performance across models (CodeBERT, CommitBART, UniXcoder, CodeT5, and GRAPE) and datasets (JavaVFC, PatchDB, and Devign) provides evidence that the cause of poor performance stems from how the data from different sources, i.e., NVD-associated patches and manually identified security patches, exhibit different characteristics.




\begin{table}[t]
\centering
\caption{Experiment result on LLMs with chain-of-thought.}
\label{tab:result-llm}
\begin{tabular}{lrrr}
\toprule
\textbf{Model} & \textbf{Precision} & \textbf{Recall} & \textbf{F1-Score} \\
\midrule
DeepSeek-R1-7B      & 53.76 & 53.48 & 53.62 \\
Llama-3.1-8B        & 55.28 & 37.84 & 44.93 \\
DeepSeek-Coder-7B   & 41.27 & 21.42 & 28.20 \\
GPT-4.1             & 54.12 &  6.04 & 10.87 \\
Qwen-8B             & 64.29 &  5.91 & 10.83 \\
GPT-4.1-mini        & 59.09 &  5.12 &  9.43 \\
o3-mini             & 81.25 &  3.42 &  6.56 \\
\midrule
Random Guess        & 49.93 & 48.75 & 49.34 \\
\bottomrule
\end{tabular}
\end{table}

Furthermore, we also conducted experiments with LLMs to see whether the pretraining knowledge helps in identifying security patches in the wild by utilizing the JavaVFC dataset.
First, we leveraged the Chain-of-Thought prompting without fine-tuning with several commercial and open-source LLMs. As a result, it is revealed that in their current state, they are no better than Random Guess, as shown in Table ~\ref{tab:result-llm}.
In this series of experiments, the best performance was achieved by DeepSeek-R1-7B, with an F1-score of 53.62\%. While this was the highest compared to the models fine-tuned with NVD data, it provides limited insight, as the score is only marginally better than a random guess, which achieved an F1-score of 49.34\%. Other LLMs, such as several OpenAI models and Qwen3, also failed to demonstrate practical performance in identifying security patches in the wild.

\begin{table}[ht]
\centering
\caption{Experiment result for LLM Evaluation based on knowledge cut-off date}
\resizebox{\textwidth}{!}{
\begin{tabular}{lllccc}
\toprule
\textbf{Model} & \textbf{Test Data (Label Distribution)} & \textbf{Commit Date} & \textbf{Prec.} & \textbf{Rec.} & \textbf{F1} \\ \midrule
\multirow{3}{*}{Llama-3.1-8B} & JavaVFC (All) & - & 55.28 & 37.84 & 44.93 \\
                              & JavaVFC (77 Pos, 449 Neg) & After Dec 2023 & 58.82 & 38.96 & 46.88 \\
                              & JavaVFC (684 Pos, 3356 Neg) & Before Dec 2023 & 54.89 & 37.72 & 44.71 \\ \midrule
\multirow{3}{*}{DeepSeek-Coder-7B} & JavaVFC (All) & - & 41.27 & 21.42 & 28.20 \\
                              & JavaVFC (587 Pos, 3111 Neg) & After Mar 2023 & 18.82 & 23.85 & 21.04 \\
                              & JavaVFC (174 Pos, 694 Neg) & Before Mar 2023 & 41.46 & 19.54 & 26.56 \\ \bottomrule
\end{tabular}
}
\label{tab:result-llm-cutoff-date}
\end{table}

To mitigate the possibility of memorization affecting our results, we conducted two additional experiments that account for the commit timestamps of security patches in the JavaVFC dataset. In these experiments, we utilized Llama-3.1-8B (knowledge cutoff: December 2023) and DeepSeek-Coder-7B (knowledge cutoff: March 2023). This selection ensures a fair comparison, as the distribution of security patches varies considerably with their commit times. Specifically, a model with a later knowledge cutoff, such as Llama-3.1-8B, may have been exposed to a larger subset of the JavaVFC dataset during pre-training than a model with an earlier cutoff, such as DeepSeek-Coder-7B. By comparing both models, we can better assess whether performance differences stem from genuine reasoning capabilities or from prior exposure to a greater number of raw data samples within the training window.

We divided all security patches in the JavaVFC dataset, along with their corresponding negative samples, according to their commit timestamps and the model’s knowledge cutoff date. We then employed Chain-of-Thought (CoT) prompting to evaluate how pre-training knowledge influences the identification of in-the-wild security patches. 

Our observations indicate that DeepSeek-Coder-7B struggled to identify these patches effectively, achieving an F1-score below 30\% even for data that may have been included in its pre-training corpus. A similar lack of performance was observed for patches committed after their knowledge cutoff date. These findings suggest that DeepSeek-Coder-7B is incapable of reliably identifying security patches, regardless of whether it was exposed to the raw commits during pre-training or not.

As for the higher-performing Llama-3.1-8B, we found no significant difference in identification performance between patches potentially seen during pre-training and those committed after the December 2023 cutoff. This suggests that, although pre-training data may shape the model's general knowledge representation, there is no significant relationship between potential exposure to raw code and the model's ability to identify in-the-wild security patches.

\begin{table}[ht]
\centering
\caption{Performance comparison between NVD and OSS testing datasets on fine-tuned LLMs}
\resizebox{\textwidth}{!}{
\begin{tabular}{lllcccc}
\toprule
\textbf{Model} & \textbf{Train Data} & \textbf{Test Data (Provenance)} & \textbf{Prec.} & \textbf{Rec.} & \textbf{F1} & \textbf{$\Delta$F1} \\ \midrule
\multirow{3}{*}{DeepSeek-Coder-7B} & \multirow{3}{*}{MoreFixes} & MoreFixes (NVD) & 73.48 & 53.03 & 61.60 & -- \\
                          &                            & JavaVFC (OSS) & 63.61 & 32.85 & 43.33 & \color{red}$-30\%$ \\ 
                          &                            & PatchDB (OSS) & 52.31 & 64.86 & 57.91 & \color{red}$-6\%$ \\ \midrule
\multirow{3}{*}{Llama-3.1-8B}    & \multirow{3}{*}{MoreFixes}& MoreFixes (NVD) & 72.02 & 54.52 & 62.06 & -- \\
                          &                            & JavaVFC (OSS) & 59.87 & 36.27 & 45.17 & \color{red}$-27\%$ \\
                          &                            & PatchDB (OSS) & 43.81 & 77.63 & 56.01 & \color{red}$-10\%$ \\ \midrule
\multirow{3}{*}{Qwen3-8B}    & \multirow{3}{*}{MoreFixes} & MoreFixes (NVD) & 74.40 & 49.87 & 59.71 & -- \\
                          &                            & JavaVFC (OSS) & 67.58 & 32.33 & 43.73 & \color{red}$-27\%$ \\ 
                          &                            & PatchDB (OSS) & 69.74 & 31.47 & 43.37 & \color{red}$-27\%$ \\ \bottomrule
\end{tabular}
}
\label{tab:result-llm-fine-tune}
\end{table}

To complement our results, we fine-tuned three prominent LLMs, namely DeepSeek-Coder-7B, Llama-3.1-8B, and Qwen3-8B, using the NVD-derived MoreFixes dataset.
We evaluated these models across three distinct test sets: 
(1) the MoreFixes test set, representing an in-distribution NVD scenario;
(2) JavaVFC, representing in-the-wild Java security patches; and
(3) PatchDB, representing in-the-wild C/C++ patches.

As shown in Table~\ref{tab:result-llm-fine-tune}, all three LLMs failed to outperform CodeBERT when fine-tuned and tested on the NVD dataset, and they achieved only comparable performance to CodeBERT on the JavaVFC dataset.
While the LLMs showed slightly higher F1-scores on PatchDB, this result may be attributed to data memorization given that PatchDB was released in 2021.

\begin{tcolorbox}[colback=blue!5!white, fonttitle=\bfseries,title=Answer to RQ1, boxrule=1.5pt]
Models trained on NVD-sourced security patches show significantly reduced performance (6\%-90\% drop in F1-score) when evaluated with the in-the-wild data.
\end{tcolorbox}

\subsection{\textit{RQ2: How do the unreported security patches differ from security patches linked from NVD?}}\label{sec:RQ2result}
In this research question, we investigate the differences between publicly disclosed security patches and undisclosed patches. 
To start, we test whether security patches listed in the MoreFixes dataset can be easily distinguished from undisclosed security patches found in the JavaVFC dataset using a lightweight machine learning model. Next, we explore which factors drive these differences. Specifically, we focus on three areas: (1) the language used in commit messages, (2) the types of vulnerabilities addressed (as indicated by CWEs), and (3) the composition of the commits themselves.

\subsubsection{Confirming that security patches from MoreFixes and JavaVFC come from different distributions}\label{sec:xgboostexp}
We employed a shallow machine learning-based text classifier, namely XGBoost~\cite{chen2016xgboost}, to determine whether commit messages from MoreFixes and JavaVFC are lexically distinguishable.
XGBoost is a scalable tree boosting system, which has been widely used for supervised learning tasks such as classification and regression~\cite{liew2021investigation, cherif2019using, ogunleye2019xgboost}. XGBoost builds models iteratively by combining multiple weak learners (decision trees) into a strong learner, minimizing errors at each step. It optimizes model performance using advanced techniques like regularization, tree pruning, and handling missing values efficiently.
Training parameters that are configurable, such as booster type, learning rate, and maximum depth of a tree, are set with default values in our experiment.
Specifically, we set the booster type to \texttt{gbtree}, the learning rate to 0.3, and the maximum tree depth to 6.

In our experiment, we employed an undersampling method to balance the dataset between security patches from the MoreFixes and JavaVFC datasets.
In total, we use 614 patches from MoreFixes and 614 patches from JavaVFC as training data.
Using the XGBoost classifier with TF-IDF as the text representation method~\cite{qi2020text, kalra2019automatic}, we achieved an accuracy of 84\%, 
indicating that even a shallow model can effectively differentiate between MoreFixes and JavaVFC based on their lexical features.
This confirms that while security patches from both sources fix vulnerabilities, they have very different characteristics, which contribute to the poor performance of security patch detectors trained on NVD security patches to identify security patches in the wild.

\begin{tcolorbox}[colback=blue!5!white, fonttitle=\bfseries, boxrule=1.5pt]
Security patches associated with NVD can be distinguished from manually identified security patches using a shallow machine learning model, i.e., XGBoost, with an accuracy of 84\%.
\end{tcolorbox}

\subsubsection{Analysis of the commit messages}
\textbf{Perplexity}~\cite{jelinek1977perplexity} is widely used in natural language processing to quantify the uncertainty or the ``surprise'' of a model when generating a sequence of text.
A {\em higher} perplexity indicates that the model is less likely to generate the given sequence, suggesting that the evaluated sequence does not align well with the training data used during the model's pretraining. 
In contrast, a {\em lower} perplexity indicates greater alignment with the model’s training distribution.
In other words, the model expects to encounter such a sequence because it resembles the data it was trained on.

In our study, we used perplexity as a measure to explore how familiar language models are with security patch datasets. For the analysis, we utilized perplexity calculation with its formula presented in Equation~\ref{eqn:perplexity}.

\begin{equation}
    PP = \exp\left(-\frac{1}{N} \sum_{i=1}^{N} \log P(w_i | w_1, \dots, w_{i-1})\right)
    \label{eqn:perplexity}
\end{equation}

where $N$ is the sequence length and $P(w_i | w_1, \dots, w_{i-1})$
is the model's probability of the next word. Perplexity has been widely used as a measure of the performance of language models. For a given model and an input text sequence, perplexity measures the likelihood that the model will generate the text sequence. In other words, lower perplexity indicates that the model finds the text more predictable, which generally suggests higher familiarity~\cite{gonen2023demystifying, xu2024chatgpt}.

In our analysis of the commit messages, we leverage perplexity to evaluate how well a code language model, specifically CodeBERT\footnote{https://huggingface.co/microsoft/codebert-base}, recognizes the language of the commit messages associated with the security patches. We chose CodeBERT for this analysis because it is pretrained on a large corpus of code and natural language text related to code, such as their documentation, making it a good fit for analyzing code changes and commit messages. A low perplexity score suggests that the commit message follows patterns CodeBERT learned during pretraining, while a high perplexity score indicates that the model is not familiar with the observed text sequence.

\begin{figure}[h!]
    \centering
    \subfigure[Comparison of perplexity for security patches from MoreFixes versus JavaVFC based on commit message.]{
        \includegraphics[width=0.45\textwidth]{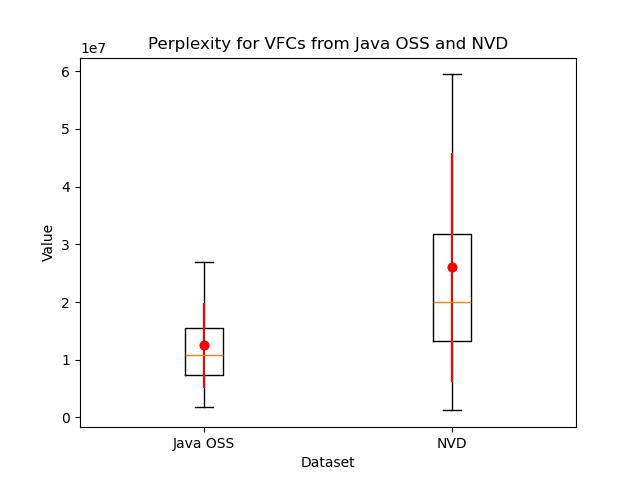}
        \label{fig:perplexity-msg}
    }
    \hfill
    \subfigure[Comparison of perplexity for security patches from MoreFixes versus JavaVFC based on commit diff.]{
        \includegraphics[width=0.45\textwidth]{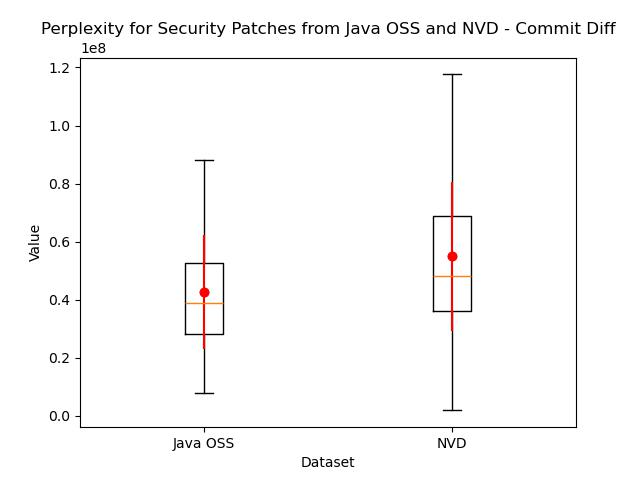}
    \label{fig:perplexity-diff}
    }
    \caption{Perplexity scores for the MoreFixes and JavaVFC datasets to measure CodeBERT's familiarity with the data.}
    \label{fig:perplexity}
\end{figure}

Figure~\ref{fig:perplexity-msg} shows the perplexity of the commit messages from the two datasets.
In the figure, the yellow line represents the median perplexity, red dot indicates the mean perplexity, and the red line represents the standard deviation.
We find that security patches from JavaVFC have a lower perplexity compared to those from MoreFixes. To determine whether this difference is statistically significant, we utilize a Mann-Whitney U test~\cite{nachar2008mann}, following guideline written by Arcuri et al.~\cite{arcuri2011practical}.

As observed from Figure~\ref{fig:perplexity}, the mean perplexity for the MoreFixes security patch commit messages is more than 1 standard deviation away from the JavaVFC commit messages.
This indicates that the commit messages in the MoreFixes security patches are less regular compared to the ones in JavaVFC, and that the MoreFixes security patches are more unique and distinct compared to the security patches found in-the-wild, such as those in the JavaVFC dataset.

On the other hand, Figure~\ref{fig:perplexity-diff} presents the perplexity scores for each dataset based on the commit diff, i.e., code changes.
In the code changes analysis, JavaVFC still presents a lower perplexity score compared to MoreFixes.
This indicates that JavaVFC, which was sourced from unreported security patches, contains patches that align more closely with CodeBERT's pretraining data.
This suggests that such unreported patches are more commonly found in the wild.
In contrast, despite MoreFixes' patches also being hosted on open source repositories, it still shows a higher perplexity score. 
This indicates that the patches in MoreFixes occur less frequently in the pretraining data compared to the patches from JavaVFC, highlighting their relatively lower prevalence in open-source code.
Using the Mann–Whitney U test, we confirmed that these differences are statistically significant (p-value < 0.001).

\begin{figure}[h!]
    \centering
    \subfigure[Comparison of perplexity for security patches from MoreFixes versus JavaVFC based on commit message (Intra-project setting)]{
        \includegraphics[width=0.45\textwidth]{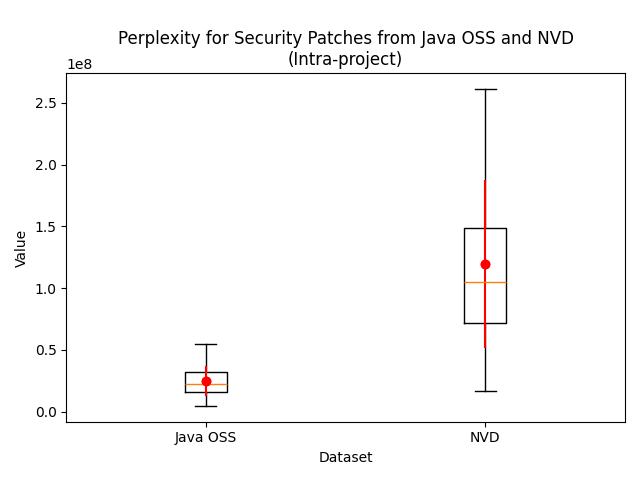}
        \label{fig:perplexity-msg-intraproject}
    }
    \hfill
    \subfigure[Comparison of perplexity for security patches from MoreFixes versus JavaVFC based on commit diff (Intra-project setting.]{
        \includegraphics[width=0.45\textwidth]{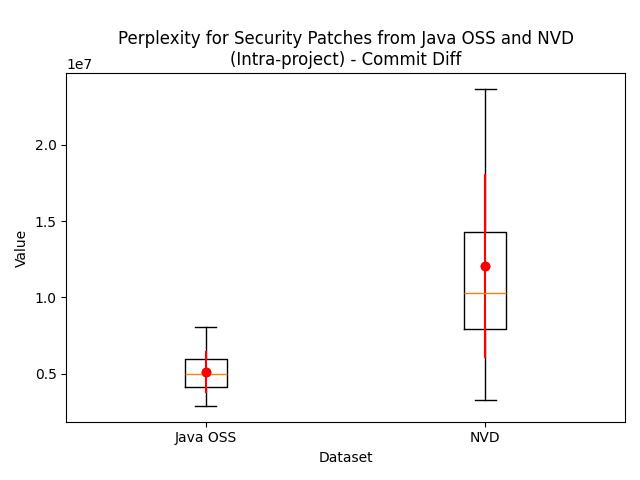}
    \label{fig:perplexity-diff-intraproject}
    }
    \caption{Perplexity scores for MoreFixes vs JavaVFC datasets in intra-project setting.}
    \label{fig:perplexity-intraproject}
\end{figure}

Additionally, we conducted a perplexity analysis on commit messages and commit diffs within the same set of projects (i.e., intra-project settings) to rule out potential bias from project characteristics or writing styles.
For the intra-project settings, we filtered the MoreFixes dataset to include only projects that appear in the JavaVFC dataset, using the same set of projects as in the experiment described in Section~\ref{sec: overlap-data-intra-project}.
As shown in Figure~\ref{fig:perplexity-intraproject}, similar observations were found in the intra-project setting. This indicates that perplexity analysis exhibits consistent behavior when used to compare the characteristics of NVD-linked and unreported security patches.

Please note that, in this experiment, we only compare  NVD-linked patches with JavaVFC patches. We do not compare NVD-linked patches with Devign patches because we utilize CodeBERT, which is not pretrained on C/C++ language. Therefore, we only compare it within the Java language to ensure our analysis is not biased toward the model's unfamiliarity with C/C++ language.

If the perplexity values for security patches from the MoreFixes and JavaVFC datasets were similar, it would suggest that these datasets share similar characteristics, as the pretrained model perceives them with a comparable level of uncertainty.
However, as shown in Figures~\ref{fig:perplexity} and~\ref{fig:perplexity-diff}, the perplexity values for JavaVFC and MoreFixes datasets differ significantly.
Specifically, the probability distribution of MoreFixes data is less frequently encountered in CodeBERT's training data, which was collected from a vast number of open-source projects.

This indicates that patches registered in the NVD are more unique compared to security patches found in the wild, such as those in the JavaVFC dataset.
Furthermore, it suggests that vulnerabilities reported to the NVD tend to focus on security issues that are less frequently encountered in real-world scenarios.
In other words, the security patches found in JavaVFC may occur more frequently in the real world but are not reported to the NVD.
Hence, introducing security patches from JavaVFC will incorporate these overlooked patches, which are not reported to the NVD, helping the model better learn to identify security patches in the real world, whether they are reported or unreported.

As an example of a security patch that is not reported to NVD, we present a case where a potential deadlock vulnerability affecting the availability of a database goes unreported, using Apache Doris as a case study.\footnote{https://github.com/apache/doris}
Apache Doris is a high-performance, real-time analytical database with an MPP (Massively Parallel Processing) architecture, optimized for sub-second query responses across both high-concurrency and complex analytical workloads.
In a particular commit\footnote{https://github.com/apache/doris/commit/6f20cac1da20561479c93}:

\fbox{%
\parbox{0.95\textwidth}{%
\textbf{Commit Message}
[fix](cooldown) Fix potential deadlock while calling \texttt{handleCooldownConf}
}
}

This commit aims to fix a vulnerability that can occur when \texttt{TableScheduler} and \texttt{tabletReport} are called simultaneously. 
When this happens, the database stops working and enters a deadlock state because a table's write and read locks are held by different operations.
This type of vulnerability is not reported to the NVD and is treated as a bug.
In reality, it could affect the availability of the program, since it can be exploited for a denial-of-service attack.

Similarly, we present an example of a security patch linked to the NVD that has highly irregular language.
This commit, taken from the libxml2 library\footnote{https://github.com/GNOME/libxml2/commit/932cc9896ab41475d4aa429c27}, has the highest perplexity value among all commits analyzed, with a score of $8.21 \times 10^8$.

\fbox{%
\parbox{0.9\textwidth}{%
\textbf{Commit Message}

Fix buffer size checks in \texttt{xmlSnprintfElementContent}. \\
\texttt{xmlSnprintfElementContent} failed to correctly check the available
buffer space in two locations. \\
Fixes bug 781333 (CVE-2017-9047) and bug 781701 (CVE-2017-9048). \\
Thanks to Marcel Böhme and Thuan Pham for the report. 
}
}

In this example, the commit message mentions that it fixes two CVEs, along with their unique identifiers. The higher perplexity score for such commit messages is expected, because the numerous unique identifiers in the sequence are not anticipated by CodeBERT, leading to greater perplexity.

In summary, our results show that unreported in-the-wild security patches from Java OSS have lower perplexity than reported patches. This suggests that unreported patches occur more frequently in the wild, limiting the ability of a language model trained only on reported security patches to identify them.

\begin{tcolorbox}[colback=blue!5!white, fonttitle=\bfseries, boxrule=1.5pt]
Security patches linked to reported vulnerabilities in the NVD exhibit less regular lexical usage compared to the unreported security patches, reflecting a bias towards usage of uncommonly used terms in the commit message. These messages often include explicit references to vulnerabilities, deviating from established best practices for commit documentation.
\end{tcolorbox}

\subsubsection{Vulnerability types} 
While we observe that the commit messages of the security patches exhibit distinctive properties, these differences do not explain the decrease in performance of the GRAPE classifier, as shown in Table \ref{tab:result-per-language}.
The GRAPE classifier relies solely on code changes and does not use the commit message.
To explore this further, we analyze the semantic content of the patches, specifically focusing on the vulnerabilities they address.
We begin by comparing the CWE distribution between security patches from MoreFixes and those from JavaVFC.
Our analysis reveals that the top-10 most frequent CWE classes in MoreFixes differ from those in the security patches sampled from the JavaVFC dataset.

To investigate the vulnerability types in the JavaVFC dataset, we manually sampled and analyzed a subset of 50 security patches, assigning each patch a corresponding CWE.
Due to the labor-intensive nature of this task, we limited our sample size to 50 patches.
Manual labeling was performed by two annotators, achieving a kappa statistic score of 0.8, indicating a substantial agreement~\cite{viera2005understanding}. 

The two human annotators first familiarized themselves with CWEs by studying the CWE tree provided on the NVD website.\footnote{https://nvd.nist.gov/vuln/categories/cwe-layout\#}
When a report is categorized under a higher-level CWE, the annotators would then visit the corresponding definition page on the CWE website and adjust their CWE ID, e.g., \url{https://cwe.mitre.org/data/definitions/119.html}.
From the definition page, the annotators would check the \texttt{Relationship} section to classify the vulnerability into a lower CWE class. This process is repeated iteratively until the vulnerability no longer fits into any lower-class CWE ID or until the depth of the label reaches the fourth level.

For the manual labelling, we first consider all of the parent classes of the assigned CWE ID in an attempt to generalize the category of vulnerabilities found in JavaVFC.
In cases of disagreement between the two annotators, they engage in discussions to reach a consensus.
A total of 5 days were spent classifying the sampled vulnerabilities from JavaVFC.

\begin{table}[t]
\centering
\caption{CWE ID distribution for NVD data.}
\label{tab:cwe-nvd-all}
\begin{tabularx}{\columnwidth}{lXr}
\toprule
\textbf{CWE-ID} & \textbf{CWE Name} & \textbf{Percentage} \\
\midrule
CWE-79        & Improper Neutralization of Input During Web Page Generation & 13.9\% \\
CWE-noinfo    & --                                                          & 10.1\% \\
NVD-CWE-Other & --                                                          & 8.4\%  \\
CWE-20        & Improper Input Validation                                   & 6.1\%  \\
CWE-119       & Improper Restriction of Operations within the Bounds of a Memory Buffer & 4.8\% \\
CWE-200       & Exposure of Sensitive Information to an Unauthorized Actor   & 4.6\%  \\
CWE-787       & Out-of-bounds Write                                         & 4.4\%  \\
\bottomrule
\end{tabularx}
\end{table}

\begin{table}[t]
\centering
\caption{CWE ID distribution for sampled JavaVFC data.}
\label{tab:cwe-java-oss}
\begin{tabular}{llr}
\toprule
\textbf{CWE-ID} & \textbf{CWE Name} & \textbf{Percentage} \\
\midrule
CWE-664 & Improper Control of a Resource Through its Lifetime       & 30\% \\
CWE-703 & Improper Check or Handling of Exceptional Conditions      & 26\% \\
CWE-691 & Insufficient Control Flow Management                      & 26\% \\
CWE-755 & Improper Handling of Exceptional Conditions                & 19\% \\
CWE-284 & Improper Access Control                                    & 16\% \\
\bottomrule
\end{tabular}
\end{table}

The CWE distribution of MoreFixes and JavaVFC datasets are presented in Tables~\ref{tab:cwe-nvd-all}  and~\ref{tab:cwe-java-oss}, respectively. 
Based on the labelling result, the top three CWEs in JavaVFC are CWE-664 (Improper Control of a Resource Through its Lifetime), CWE-703 (Improper Check or Handling of Exceptional Conditions), and CWE-691 (Insufficient Control Flow Management) vulnerabilities.
These types of vulnerabilities are closely associated with flaws in the implementation phase, as stated in the CWE's \texttt{Modes of Introduction} section. 
Additionally, they often result in denial of service or loss of availability.

On the other hand, the CWE distribution in the MoreFixes dataset shows a more even spread of CWEs. 
However, approximately 18.5\% of the data in MoreFixes is not explicitly labelled as it contains \texttt{CWE-noinfo} and \texttt{NVD-CWE-Other label}.
This suggests that additional effort is needed to populate vulnerability data for a high-quality and complete vulnerability database.
The top two CWE classes that appear in NVD are related to the user input, which are CWE-79 
(Improper Neutralization of Input During Web Page Generation) and CWE-20 
(Improper Input Validation).
When contrasting Table~\ref{tab:cwe-nvd-all} and Table~\ref{tab:cwe-java-oss}, there is a notably different CWE distribution between MoreFixes and JavaVFC datasets.
None of the top 5 CWEs registered in MoreFixes overlap with the top 5 CWEs sampled from JavaVFC. 
Security patches related to XSS or improper validation are more likely to be reported, while unreported security patches tend to be related to exception or resource handling.
This demonstrates the existence of data drift from a CWE-type perspective, meaning that the training and testing data follow different CWE type distributions. As suggested by other studies, such performance drops will continue to occur if this issue is not properly addressed~\cite{lu2018learning, mallick2022matchmaker}.

To check whether there is a tendency for why certain CWE-IDs are reported to NVD, we analyzed the top CWE IDs found in MoreFixes and sampled JavaVFC dataset against the top 25 most dangerous software weaknesses~\cite {top25dangerousvuln}.
We found that the common CWE IDs in the MoreFixes dataset are included in the top 25 most dangerous list, while none of the CWE IDs from the sampled JavaVFC dataset appear there.
This suggests that developers might be biased toward reporting only vulnerabilities they consider easier to identify and exploit. 
This finding is consistent with prior work~\cite{alexopoulos2021vulnerability} that showed low levels of participation in the vulnerability reporting process even for widely used projects.
Nevertheless, it remains crucial to report all security weaknesses that could potentially affect a system’s availability, integrity, or confidentiality. 

In addition to limited awareness of best practices for reporting vulnerabilities~\cite{braz2022less, thomas2018security}, underreporting may also stem from the fact that developers often receive greater monetary incentives when submitting reports directly to the application developers or through bug bounty programs such as HackerOne~\footnote{https://hackerone.com/bug-bounty-programs}. Furthermore, when developers discover vulnerabilities in their own applications, they may be disincentivized to report them, especially if doing so would reveal vulnerabilities they themselves introduced~\cite{markcurphey}.

To investigate further whether the distribution of CWE types affect the model's performance greatly, we conducted an experiment with a similar distribution of CWE types as training data. Please note that the selected CWE types are based on the manually labeled data sampled from JavaVFC. We extracted the top five CWE classes from MoreFixes and conducted an experiment with CodeBERT, using a total of 288 security patches from CWE-664, CWE-703, CWE-691, CWE-755, and CWE-284 as training data. The results are presented in Table~\ref{tab:overlap-cwe-result}.

However, the result shows that CodeBERT failed to learn from the limited data, predicting all data points as security patches.
This outcome underscores a key limitation of the current datasets, which is the absence of fine-grained vulnerability type labels for in-the-wild patches. Although datasets such as JavaVFC, PatchDB, and Devign provide valuable collections of unreported patches, they do not include explicit CWE classifications or detailed vulnerability type annotations. This lack of labeling constrains the ability to perform more granular analyses, such as evaluating model performance across specific vulnerability categories. Addressing this limitation would require additional manual labeling or the creation of new benchmark datasets that systematically categorize in-the-wild patches.


\begin{table}[t]
\centering
\caption{Result from using data from the top 5 CWE types in sampled JavaVFC as training data.}
\label{tab:overlap-cwe-result}
\begin{tabular}{llrrr}
\toprule
\textbf{Train Data} & \textbf{Test Data} & \textbf{Precision} & \textbf{Recall} & \textbf{F1} \\
\midrule
MoreFixes & MoreFixes & 50.00 & 100.00 & 66.67 \\
MoreFixes & JavaVFC   & 50.00 & 100.00 & 66.67 \\
\bottomrule
\end{tabular}
\end{table}

\begin{tcolorbox}[colback=blue!5!white, fonttitle=\bfseries, boxrule=1.5pt]
Reported vulnerabilities that are registered in NVD have a different CWE class distribution compared to the unreported vulnerabilities collected from open-source Java repositories in GitHub (represented by JavaVFC).
\end{tcolorbox}

\subsubsection{Overlap data and intra-project experiment}
\label{sec: overlap-data-intra-project}
In addition to differing vulnerability types, we further examine the overlapping data points between the NVD dataset and the JavaVFC dataset. There are 83 overlapping projects between NVD and JavaVFC, but only 15 commits appear in both datasets. This suggests that while many projects contribute to reported security vulnerabilities, a large proportion of security patches could be categorized as unreported. Therefore, this emphasizes the need to identify security patches in the wild without relying solely on reported vulnerabilities from security databases such as NVD, due to the varying characteristics between reported and in-the-wild security patches.

To mitigate the impact of project-specific factors in our results, we conducted an experiment in which we trained and tested the model within certain projects. As mentioned earlier, there are 83 projects that appear in both the NVD and JavaVFC datasets. To ensure sufficient training data for the model to learn, we filtered the projects to include only those with at least five data points. This resulted in 32 overlapping repositories between the MoreFixes and JavaVFC datasets.

Subsequently, we then trained CodeBERT on reported security patches from these 32 projects in the MoreFixes dataset and tested it on unreported security patches from the JavaVFC dataset. Due to the limited number of patches in each project, it is impractical to train on each project individually. Therefore, we combined all data points from the 32 projects. 
Furthermore, we applied the same split strategy as in the main experiment to avoid bias towards specific data points. Each project was filtered into the appropriate split, and we still obtained a relatively balanced dataset.
In total, the training set contained 587 security patches and 581 non-security patches from MoreFixes, while the test set contained 164 security patches and 164 non-security patches from JavaVFC.

The result could be seen in Table~\ref{tab:result-intraproject}. Based on the results shown, we can see that the F1-score still drops more than 13\% even when trained and tested in the same set of projects.
Additional experiments were also conducted for the C/C++ language. In this scenario, we utilized the Devign (partial) dataset consisting of two projects to better investigate the impact of intraproject-specific factors, as the dataset contains only these two projects. The results show that even when a specific project has more training data, with as many as 361 security patches from Qemu and 295 from FFmpeg, the model still struggles to detect security patches in the wild, as indicated by an F1-score drop of more than 30\%.


\begin{table}[t]
\centering
\caption{Experiment result on intraproject setting.}
\label{tab:result-intraproject}
\begin{tabularx}{\columnwidth}{llXlrrr}
\toprule
\textbf{Model} & \textbf{Train Data} & \textbf{Provenance} & \textbf{Test Data} & \textbf{Prec.} & \textbf{Rec.} & \textbf{F1} \\
\midrule
CodeBERT & MoreFixes (Java)  & NVD       & MoreFixes (Java)  & 71.43 & 64.10  & 67.57 \\
CodeBERT & MoreFixes (Java)  & Java OSS  & JavaVFC           & 58.90 & 58.54  & 58.72 \\
CodeBERT & MoreFixes (C/C++) & NVD       & MoreFixes (C/C++) & 81.82 & 100.00 & 90.00 \\
CodeBERT & MoreFixes (C/C++) & C/C++ OSS & Devign            & 52.68 & 69.73  & 60.02 \\
\bottomrule
\end{tabularx}
\end{table}

\subsubsection{Composition of the commits}
Apart from the language used in the commit messages and the types of vulnerabilities fixed by the security patches, we observed that the NVD dataset contains several sources of noise, such as meta-changes and non-code changes. 
This suggests that the commits linked to NVD vulnerability reports are not always security patches, even though the quality of the links was previously assumed when constructing the dataset of security patches. 

\textbf{Meta-change commits.} First,  we found that  NVD reports occasionally link to meta-changes, such as merge commits.
Merge commits are those with two parent commits in the revision history, typically created when integrating changes from one branch (e.g., a feature branch) into another (e.g., the main or release branch).
While the security fix may be included within a subset of changes made in these commits, there is often a more granular commit on one of the branches that should have been identified instead. 
While these meta-change commits still provide valuable information and should be part of the vulnerability record, they may be less useful for building a training dataset aimed at detecting vulnerability fixes. 
In the MoreFixes dataset, merge commits account for 12.03\%, while in JavaVFC, they make up 6.96\%.

\begin{figure*}[]
    \centering
    \includegraphics[width=1\textwidth]{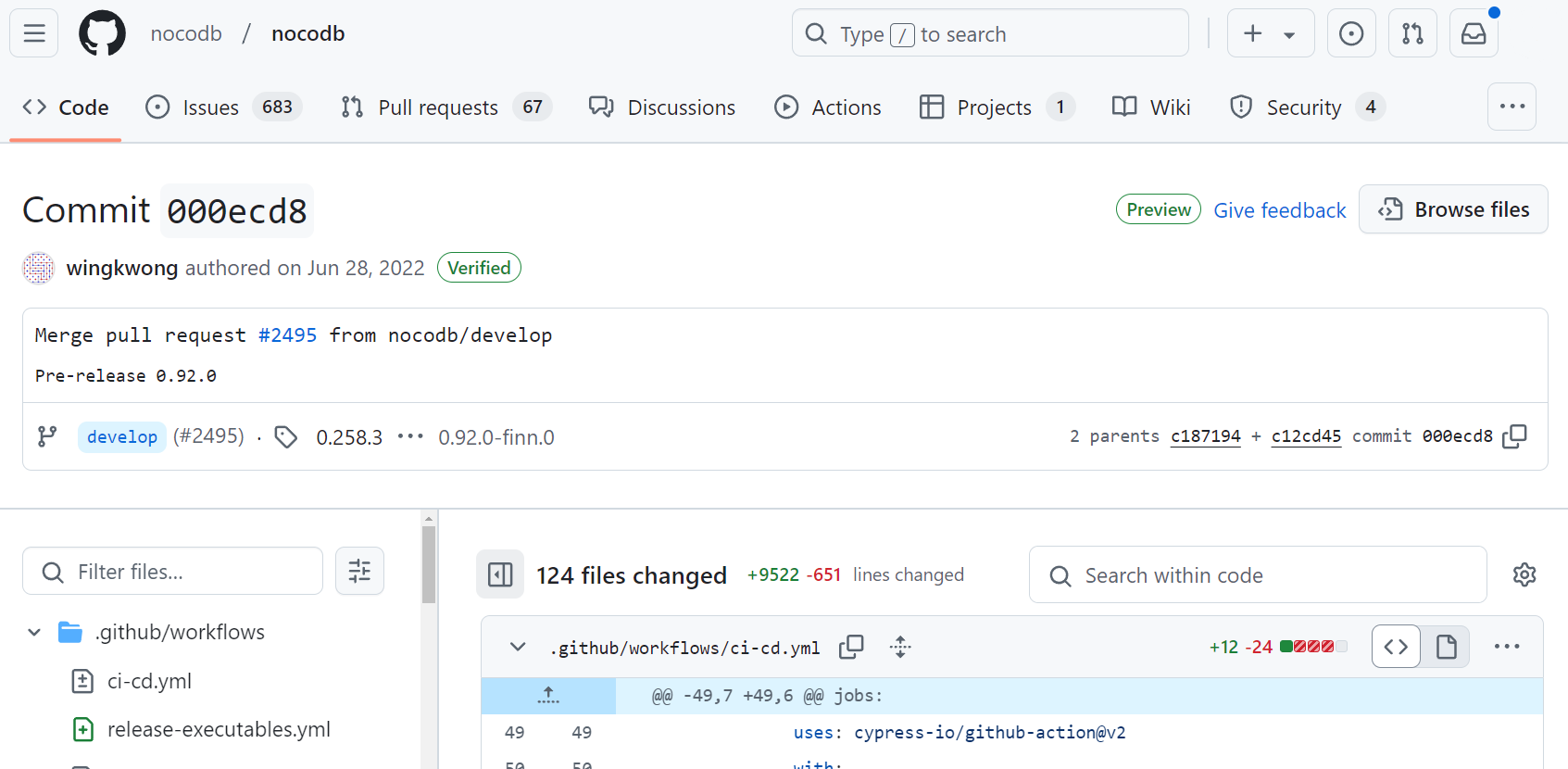}
    \caption{Example of a merge commit linked from NVD that integrates the \texttt{develop} branch into a \texttt{release} branch. This commit changes over a hundred files. The actual security fix (made in a single commit https://github.com/nocodb/nocodb/pull/2495/commits/ac346945f6cd4d1e371d57267d80f6dfdbbcc605) is a small subset of all changes (composed of over 180 commits) made in the merge commit. 
    }
    \label{fig:example_nocodb}
\end{figure*}

\begin{figure*}[]
    \centering
    \includegraphics[width=0.85\textwidth]{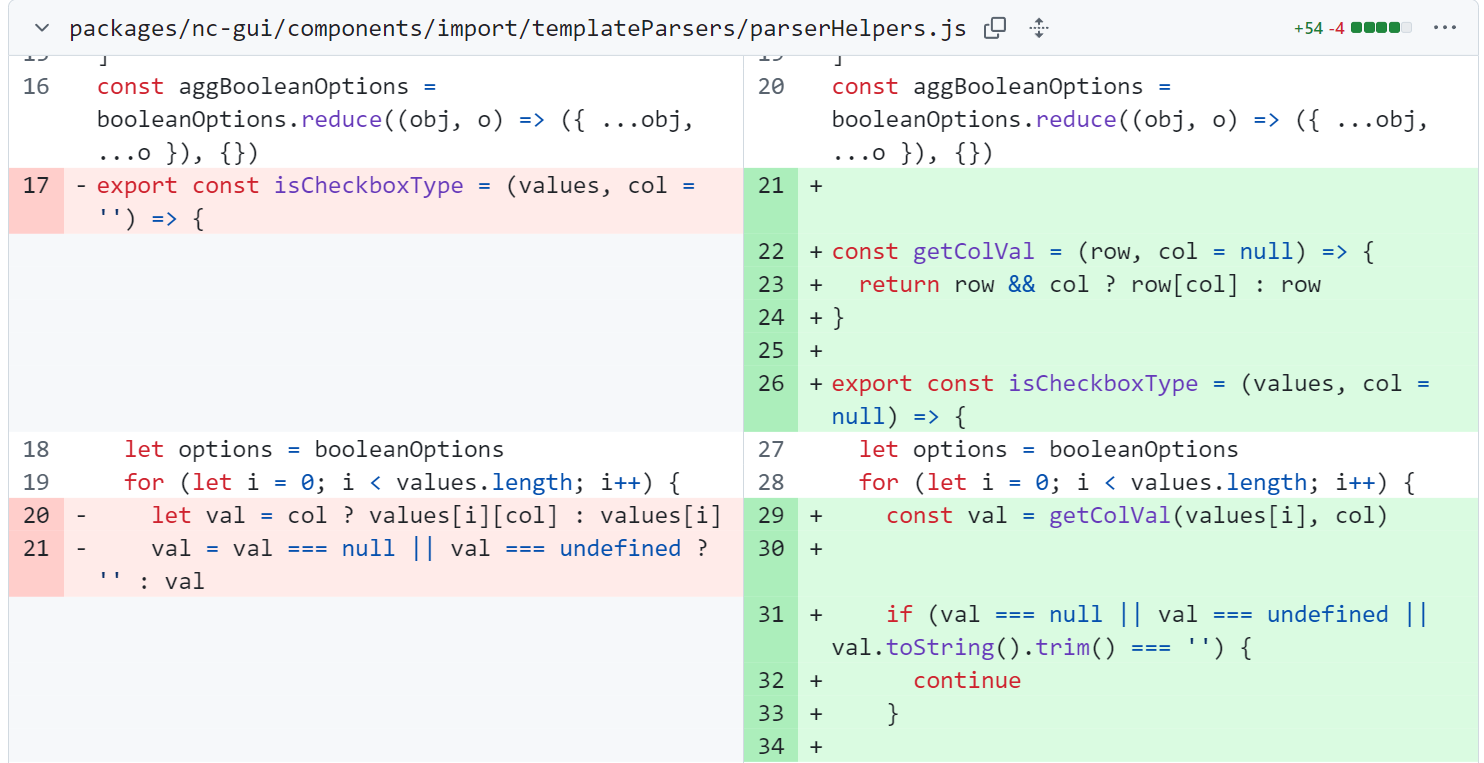}
    \caption{Example of a code change made to implement a feature, unrelated to the vulnerability fix. 
    }
    \label{fig:parserhelper}
\end{figure*}

Merge commits often encompass multiple modifications, changing up to several hundred files.
An example of such a merge commit is given in Figure \ref{fig:example_nocodb}, obtained from the NocoDB project\footnote{\url{https://github.com/nocodb/nocodb/commit/000ecd886738b965b5997cd905825e3244f48b95}}. 
This commit is linked to the NVD as the security patch for CVE-2022-3423, but it involves changes to several hundred files. 
The vulnerability is classified under CWE-770 (Allocation of Resources Without Limits or Throttling) and has a medium severity.
Over 20\% of the changes are non-code related, including at least 26 configuration or documentation files (identified by their file extensions: (\texttt{.yml}, \texttt{.md}, and \texttt{.json}).
For the purpose of training a security patch detector, this commit does not provide useful information.
Figure \ref{fig:parserhelper} illustrates part of the changes in the NocoDB commit, showing a code change for refactoring code that is unrelated to the vulnerability.

\textbf{Non-code changes.}
For example, some patches contain non-vulnerability-fixing changes, such as documentation updates or test cases.
In contrast, the JavaVFC dataset does not include any security patches that consist solely of test cases or documentation updates.
All changes in the JavaVFC dataset involve at least one modification to the code files.
On the other hand, 1\% of commits in the MoreFixes dataset contain only documentation changes (identified by file extensions such as ``.md'' and ``.txt'') or updates to test files related to the actual fix (identified through a case-insensitive search for filenames containing ``test'').

To further analyze the proportion of code changes that did not affect the executions of the code, we examined the cyclomatic complexity of the code modified by the security patch.
The cyclomatic complexity for security patches in the MoreFixes dataset was obtained from the  CVEFixes~\cite{bhandari2021cvefixes} data.
Our analysis reveals that 12\% of the security patches in the NVD dataset do not have a valid complexity score (indicated as 0 in the MoreFixes dataset), suggesting that these security patches did not modify executable code.
Instead, they likely modify files related to configuration or markup (e.g., json, .env, HTML). 
In the MoreFixes dataset, approximately 1\% of commits only modify test cases or documentation, whereas the JavaVFC dataset does not contain commits that solely change test cases or documentation.


\begin{table}[t]
\centering
\caption{Experimental results of removing noise in NVD dataset.}
\label{tab:rq4-remove-noise}
\begin{tabular}{llrrr}
\toprule
\textbf{Train Data} & \textbf{Test Data} & \textbf{Precision} & \textbf{Recall} & \textbf{F1-Score} \\
\midrule
MoreFixes (w/ noise)  & MoreFixes & 61.60 & 83.72 & 70.97 \\
MoreFixes (w/o noise) & MoreFixes & 62.84 & 69.96 & 66.21 \\
\midrule
MoreFixes (w/ noise)  & JavaVFC   & 50.00 & 63.16 & 55.81 \\
MoreFixes (w/o noise) & JavaVFC   & 61.76 & 55.26 & 58.33 \\
\bottomrule
\end{tabular}
\end{table}

To account for the aforementioned factors, such as meta-changes and non-code changes, that negatively impact the model's ability to identify security patches in the wild, we conducted additional data cleaning. Specifically, we constructed a training dataset consisting only of security patches that modify a single file per commit and excluded documentation or metadata files such as JSON, XML, TXT, or configuration files. This cleaning process aimed to address issues caused by multiple changes within a single commit, which often introduce irrelevant modifications from meta-changes, as well as to evaluate the impact of removing non-code changes that typically do not contribute directly to the patch.
The results are presented in Table~\ref{tab:rq4-remove-noise}. After removing noisy files with potentially irrelevant changes, we observed an improvement in identifying in-the-wild patches. When trained on the full dataset, the model achieved an F1-score of 55.81\%, whereas after noise removal, the score increased by 4.5\% to 58.33\%.

On the other hand, a slight performance degradation was observed on the NVD dataset, with the F1-score dropping from 70.97\% to 66.21\%. 
This provides further evidence that improving the performance of security patch identification in the wild may lead to lower performance on benchmarks composed of NVD-linked security patches. 
Therefore, the current practice of benchmarking security patch identifiers solely on reported security patches may not necessarily translate into improved performance on in-the-wild patches.

Overall, these findings highlight the presence of data drift, which causes models trained on NVD data to experience performance drops when applied to identify in-the-wild security patches~\cite{lu2018learning, rahmani2023assessing, mallick2022matchmaker}.
This underscores the importance of constructing clean, high-quality datasets that accurately represent in-the-wild security patches, in order to develop reliable patch identification models that extend beyond theoretical research.

\begin{tcolorbox}[colback=blue!5!white, fonttitle=\bfseries,title=Answer to RQ2, boxrule=1.5pt]
The security patches associated with reported vulnerabilities in MoreFixes and those collected from JavaVFC can be distinguished from each other with a high accuracy (close to  85\%).
The differences stem from the writing of the commit messages, 
distributions of the vulnerability types, and composition of the commits. 
\end{tcolorbox}

\subsection{\textit{RQ3: Does combining data from both sources, NVD and Java OSS data, lead to better and more robust performance?}}
As our analysis in Section~\ref{sec:RQ1results} shows, the NVD data and the in-the-wild samples essentially assess two different aspects of security patch detectors.
To be precise, the NVD data focuses on known vulnerabilities that have been formally reported and well-documented.
In contrast, in-the-wild samples come from real-world, open-source repositories, such as GitHub, including patches that fix unreported security vulnerabilities or subtle security issues.

To address the loss of efficacy observed in the in-the-wild experiments, we construct a new dataset that combines data from both sources. 
Specifically, this new dataset includes the MoreFixes dataset together with the in-the-wild security patch data.

First, we split the JavaVFC dataset into an 8:1:1 ratio for training, validation, and testing, sorted by commit timestamps.
Next, we incorporated the training and validation splits from JavaVFC into the MoreFixes data to create combined training and validation splits. 
Negative samples for the experiment were collected by sampling a random commit from the same repository at a ratio of 1:1. Thus, for each security patch, we obtained a random commit to be used as non-security patch data.
Finally, we evaluated the performance of the model trained on the combined data using the test split from the JavaVFC dataset. 
Incorporating data from JavaVFC into the MoreFixes dataset improves the model’s performance in detecting security patches in the wild.

Table~\ref{tab: rq3} presents the results of this experiment. 
When the CodeBERT model is trained solely on MoreFixes data and tested on JavaVFC data, it achieves an F1-score of 55.81\%. 
Incorporating additional Java OSS data during training raises the F1-score to 77.99\%, an improvement of over 20\%. 
Moreover, when evaluated on the NVD test set, the model trained with both MoreFixes and JavaVFC data performs on par with the model trained on MoreFixes data alone, with less than a 1\% difference in the F1-score. 
This shows that training models using security patches associated with NVD and data of unreported vulnerabilities is sufficient to prevent their poor performance on in-the-wild security patches.

While constructing the combined dataset by adding all training data from an NVD dataset (MoreFixes) and JavaVFC leads to good performance, in practice, the JavaVFC data was constructed through a significant amount of human analysis effort.
Thus, we performed additional experiments to understand the effect of the size of the manually analyzed dataset on identifying wild-security patches. 
These experiments involved incrementally increasing the number of JavaVFC commits included in the training data. Specifically, we evaluated the performance achieved by using 100, 300, 500, and all samples of JavaVFC data (i.e., 627). 
We selected these sample sizes to investigate how incremental increases affect the introduction of curated unreported security patches. 

Table~\ref{tab:rq3-ablation} presents our results.
The performance of the CodeBERT model linearly increases based on the number of security patches added from the JavaVFC dataset.
Just by introducing 100 patches, or equivalent to less than 0.5\% of the total training data, it boosted the performance of the security patch identifier by 6 absolute points of F1-score, or 15\% increase compared to only using NVD data in the training phase.
Moreover, the performance continued to improve as more manually curated data were added, suggesting that even though these data are expensive to obtain, they can positively impact the robustness of the security patches identifier model.


\begin{table}[t]
\centering
\caption{Experimental results of constructing a new dataset that combines security patches from MoreFixes and JavaVFC in identifying in-the-wild patches with CodeBERT.}
\label{tab: rq3}
\begin{tabular}{llrrr}
\toprule
\textbf{Train Data} & \textbf{Test Data} & \textbf{Precision} & \textbf{Recall} & \textbf{F1-Score} \\
\midrule
MoreFixes           & MoreFixes & 61.60 & 83.72 & 70.97 \\
MoreFixes           & JavaVFC   & 50.00 & 63.16 & 55.81 \\
\midrule
MoreFixes + JavaVFC & JavaVFC   & 74.70 & 81.58 & 77.99 \\
MoreFixes + JavaVFC & MoreFixes & 65.40 & 75.30 & 70.00 \\
\bottomrule
\end{tabular}
\end{table}


\begin{table}[t]
\centering
\caption{Result from combining security patches from MoreFixes and JavaVFC in identifying in-the-wild security patches -- different number of JavaVFC introduced.}
\label{tab:rq3-ablation}
\begin{tabular}{lrrr}
\toprule
\textbf{\# of JavaVFC} & \textbf{Precision} & \textbf{Recall} & \textbf{F1-Score}  \\
\midrule
0 (original) & 44.39 & 44.15 & 44.27 \\
100          & 82.35 & 36.84 & 50.91 \\
300          & 72.22 & 51.32 & 60.00 \\
500          & 70.77 & 60.53 & 65.25 \\
All          & 74.70 & 81.58 & 77.99 \\
\bottomrule
\end{tabular}
\end{table}

\begin{tcolorbox}[colback=blue!5!white, fonttitle=\bfseries,title=Answer to RQ3, boxrule=1.5pt]
Training models using a mix of data from MoreFixes and JavaVFC results in robust models that achieve good performance on both classifying security patches originating from NVD and Java OSS.
Moreover, including just 100 commits from JavaVFC (i.e., less than 0.5\% training data) is enough to boost the model's F1 by 15\%.
\end{tcolorbox}

\section{Discussion}
\label{sec:discussion}

\begin{figure}[]
    \centering
    \includegraphics[width=0.8\textwidth]{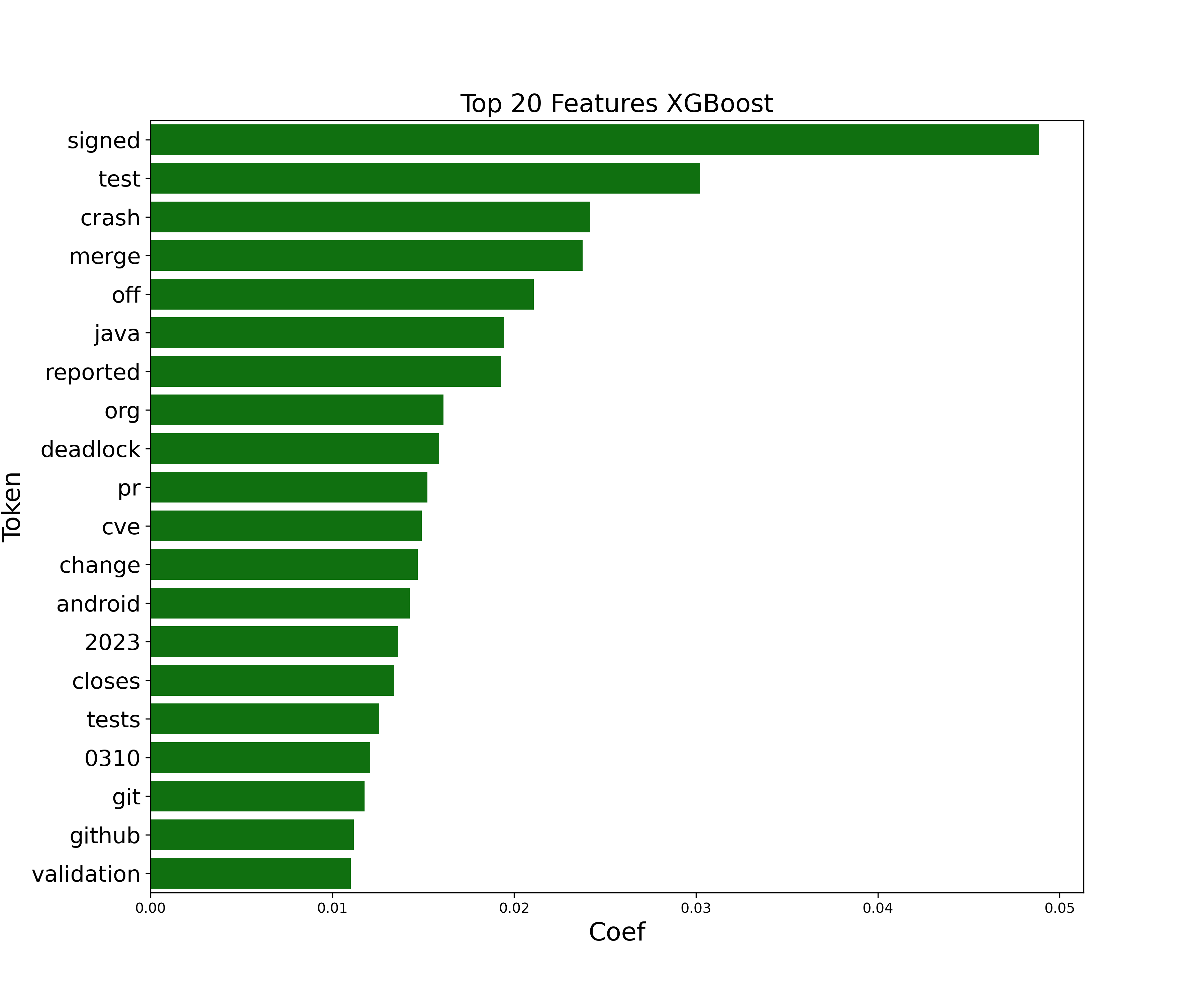}
    \caption{Top 20 most important tokens for the XGBoost classifier to distinguish between the security patches from NVD and manually identified security patches. Most of the features show that the classifier relies on information unrelated to the actual contents of the code changes. The classifier learns correlations based on meta-information, such as whether a commit is a merge commit, has been officially signed-off by a maintainer, or explicitly includes a CVE ID, which are more commonly observed in NVD-associated commits. }
    \label{fig:xgb-features}
\end{figure}

\subsection{NVD-associated Security Patches Are Rarely Silent} 
Our analysis of the language used in the commit messages of NVD-associated patches suggests that they do not resemble security patches frequently made during the software development process. 
In Section \ref{sec:xgboostexp}, we present an analysis using XGBoost to distinguish security patches found in the wild and those listed in NVD. 
XGBoost enables us to interpret the features that it considers the most important for making predictions or decisions within the model. 
For a single decision tree, feature importance is determined by the extent to which each attribute’s split improves the performance measure (e.g., Gini index), weighted by the number of observations. These values are aggregated to reflect the total impurity reduction contributed by each feature and averaged across all decision trees in the XGBoost model.
Using the features identified as important by XGBoost, we perform a deeper analysis to derive further insights.

Figure~\ref{fig:xgb-features} presents the most important features identified by XGBoost.   
We observe that a significant proportion of these important features are tokens related to the metadata of the commits, e.g. ``merge'' occurs in merge commits, ``reported'' reveals information about the person who reported the bug/vulnerability, and ``signed'' is usually a part of the phrase ``signed off by'', indicating the developer who fixed the vulnerability. 
Additionally, ``cve'' is also identified as an important token. 
Upon further analysis, we find that 15\% of the commits in the MoreFixes dataset explicitly mention the CVE ID of the vulnerability being fixed.
This shows that a significant proportion of the security patches linked from NVD are made with direct attribution to the vulnerability they address. 
Since not all NVD vulnerability reports include links to their corresponding security patch (only 63.85\% reports contain direct links), this may suggest a bias among researchers populating the vulnerability database towards linking vulnerability reports to commits that explicitly describe the vulnerability. 
In other cases, the vulnerabilities may already be publicly known, such as when a known exploit exists or malicious actors have begun targeting the vulnerability, before the library developers implement the fix. In such scenarios, the developers may see no need to obscure the nature of the fix.


\begin{table}[t]
\centering
\caption{Experimental results comparing the effect of different modalities on model effectiveness. Training CodeBERT using only code changes does not improve robustness, leading to poor performance on in-the-wild data.}
\label{tab:result-modality}
\begin{tabular}{llllrrrc}

\toprule
\textbf{Model} & \textbf{Train Data} & \textbf{Test Data} & \textbf{Modality} & \textbf{Prec.} & \textbf{Rec.} & \textbf{F1} & \textbf{$\Delta$F1} \\
\midrule
CodeBERT & ColeFunda & ColeFunda & Code Change only
  & 98.82 & 84.77 & 91.26 & \\
CodeBERT & ColeFunda & JavaVFC   & Code Change only
  & 40.22 &  4.86 &  8.68
  & \color{red}{\textbf{--82.58}} \\
\midrule
\rowcolor[HTML]{FFFFC7}
CodeBERT & ColeFunda & ColeFunda & Msg + Code Change
  & 92.42 & 92.89 & 92.66 & \\
\rowcolor[HTML]{FFFFC7}
CodeBERT & ColeFunda & JavaVFC   & Msg + Code Change
  & 77.06 & 11.04 & 19.79
  & \color{red}{\textbf{--72.87}} \\
\bottomrule
\end{tabular}
\end{table}

\subsection {Modality of Data}
We assess the impact of the different modalities used by the models.
While commits contain information in both the commit message and the code change, prior work~\cite{zhou2021finding, han2024learning, zhou2023colefunda} suggests that only the code change should be used for training models designed to detect undisclosed vulnerabilities, i.e., silent fixes.
This recommendation arises because commit messages cannot be considered a reliable source of information in usage scenarios involving coordinated disclosure, where developers deliberately obscure their fixes.
To investigate this, we conducted an additional experiment on the CodeBERT model trained on the ColeFunda dataset, which achieved the highest performance in the previous experiment (see Section~\ref{sec:RQ1results}). 
We evaluated the model under two configurations: (1) training and testing the model with only code changes, and (2) training and testing the model with both commit messages and code changes.
As shown in Table~\ref{tab:result-modality}, incorporating commit messages during training and testing improves performance on both ColeFunda and JavaVFC evaluations. 
When tested on the same dataset (ColeFunda), the increase in F1-score is relatively low (i.e., improvement of 1.53\% over using only code changes). 
Our experiment highlights that models trained with both commit messages and code changes, as well as those trained with only code changes, exhibit substantial decreases in effectiveness when tested on the JavaVFC dataset.
While the model trained with both modalities experienced a significant F1-score drop of 72.87\% when applied in-the-wild, this decrease was less severe compared to the 82.58\% drop observed when relying solely on code changes.

\subsection{Implications and Lessons Learned}

\textbf{Security patch detectors should be assessed with commits found in-the-wild.} 
Our study underscores the importance of evaluating proposed techniques in practical, real-world settings, which is in line with prior studies~\cite{chakraborty2024revisiting, ancha2024utilizing, bui2024apr4vul}.
While prior studies focused on security patches linked from NVD, our findings suggest that research on security patch detectors should extend evaluations to include data beyond those linked from the NVD.
Our analysis reveals that a significant proportion of security patches linked from NVD were not created silently, whereas the JavaVFC dataset contains unreported vulnerabilities.
These two data sources exhibit complementary characteristics, representing different distributions of vulnerabilities. 
Specifically, the JavaVFC dataset reflects vulnerabilities with frequencies typical of those encountered during everyday software development, while NVD-linked data includes high-profile vulnerabilities.
Given these distinctions, we recommend using the JavaVFC dataset to evaluate the performance of security patch detectors.

\vspace{4px}
\noindent\textbf{Choosing a larger dataset with more programming languages may exchange model effectiveness for increased robustness. }
As shown in Table \ref{tab:result-per-language}, both the CodeBERT and GRAPE models achieved higher F1-scores on their corresponding NVD Java-only datasets, ColeFunda and GRAPE, compared to CodeBERT trained on the MoreFixes dataset, which contains commits in multiple programming languages.
When trained and tested on the ColeFunda dataset, CodeBERT achieved an F1-score of 91.26\%.
When trained and tested with its original dataset (GRAPE-data), GRAPE achieved F1-score of 79.31\%. 
In contrast, a CodeBERT model trained and tested on the MoreFixes dataset, containing multiple programming languages, achieved an F1-score of only 70.97\%.
These results suggest that using a dataset with commits exclusively from the Java programming language can lead to more effective models. 
However, this increased effectiveness comes at the cost of reduced robustness when applied to commits from different datasets, such as the JavaVFC dataset. CodeBERT and GRAPE models, when trained on Java-only datasets, showed 45-90\% drop in F1-score when tested on security patches from JavaVFC repositories.
In comparison, models trained on the MoreFixes dataset exhibited only a 30\% performance drop under the same conditions.
As noted earlier, the MoreFixes dataset is substantially larger than the Java-only datasets, which may contribute to its greater robustness across different testing scenarios.

\vspace{4px}
\noindent\textbf{Training on only code changes does not sufficiently improve models' robustness for detecting unreported vulnerabilities.}
Prior studies~\cite{zhou2021finding,nguyen2023multi,zhou2023colefunda} suggested that models for detecting security patches should be trained solely on the code change information in commits to ensure robustness against the lack of vulnerability information in commit messages for silent fixes. However, our experiments indicate that this approach is insufficient.
Our findings show that training models exclusively on code changes from NVD-linked security patches does not prevent a significant performance drop when deployed on commits from the JavaVFC dataset, with the F1-score decreasing by 82\%, as shown in Table~\ref{tab:result-modality}.
In contrast, incorporating both commit messages and code changes reduces the performance drop to 72\%. While still suboptimal, this result demonstrates an improvement over using code changes alone. These findings suggest that relying solely on code changes does not enhance robustness when applied to security fixes addressing vulnerabilities that were not publicly disclosed.

\vspace{4px}
\noindent\textbf{NVD-linked data can contain noise.} 
Similar to other studies~\cite{croft2023data} that have identified issues with data cleanliness or bias in datasets used for software engineering research, we find that security patches linked from NVD can contain noise originating from several sources.  
For example, NVD may link vulnerabilities to commits that do not involve code changes, or to commits that include a superset of changes beyond those made to fix the vulnerability. 
This highlights the importance of applying data cleaning methods~\cite{northcutt2021confident} to improve the quality of NVD-linked data. Additionally, future research could benefit from incorporating diverse data sources, as we have done in this study, to reduce overfitting and enhance the robustness of models trained on such datasets.

\vspace{4px}
\noindent\textbf{Building more robust vulnerability datasets by incorporating patches found in-the-wild.} Detecting security patches is one approach used by several studies~\cite{zhou2021finding, zhou2023colefunda, han2024learning, sawadogo2020learning, sawadogo2022sspcatcher}, to construct datasets for evaluating vulnerability detection approaches.
These datasets are often created based on commits linked from NVD vulnerability reports. 
However, our study suggests that such datasets represent only a small, skewed subset of all security patches. In the future, we plan to analyze how the selection of commits used in constructing vulnerability datasets may influence the effectiveness of vulnerability detection models.

\subsection{Threats to Validity}
\label{sec:threats}
\vspace{0.2cm}\noindent{\bf Threats to Internal Validity.} Threats to internal validity refer to potential errors in our experiments and other implementation issues. 
In this study, we consider multiple approaches. 
We successfully replicated the strong, state-of-the-art performance of GRAPE and CodeBERT from prior studies~\cite{han2024learning, sun2023silent}.
Although ColeFunda did not have a publicly available implementation, we were able to train a CodeBERT model that achieved equally strong performance, with an AUC of 0.9, comparable to or higher than the reported AUC of 0.8~\cite{zhou2023colefunda}, using the publicly released ColeFunda dataset. 
We also thoroughly validated our implementation, and the strong performance of the multiple models was consistently replicated across various datasets constructed from the commits linked to the NVD.

Additionally, while our use of random sampling for unreported GitHub commits may introduce some noise in the negative labels, we expect its impact to be minimal. This is supported by previous studies~\cite{wang2019detecting, li2017large, li2018vuldeepecker}, which highlight that the proportion of reported security patches in the wild is less than 0.4\% of the total commit population, and under 10\% even among randomly selected commits that were manually checked. Given the low prevalence of unreported security patches and our sampling strategy, we believe that any potential noise has a negligible effect on our experimental results.

\vspace{0.2cm}\noindent{\bf Threats to Construct Validity.} 
A potential threat to construct validity in our study concerns the choice of evaluation metrics.
We employed precision, recall, and F1-score as our evaluation metrics, which are well-established and widely used in classification tasks, as demonstrated in previous studies\cite{han2024learning, xu2019sentiment, zhang2020sentiment, wu2022enhancing, zhou2021spi}.
More importantly, these metrics are particularly relevant to our context, as detecting silent security patches involves identifying a small number of true positives within a large set of non-security commits.
Precision and recall directly capture the trade-off between detecting as many true security patches as possible while minimizing false alarms, whereas the F1-score provides a balanced measure of both.
Therefore, while these metrics are standard in classification tasks, their validity in this study lies in their ability to accurately reflect the model’s effectiveness in detecting silent security patches, thereby minimizing the threat to construct validity.

\vspace{0.2cm}\noindent{\bf Threats to External Validity.} 
Our study utilizes the JavaVFC, PatchDB, and Devign datasets, which consist of security patches from Java and C/C++ projects. 
This presents a potential threat to external validity, as the results may not directly generalize to other programming languages.
However,  this threat is mitigated by the inclusion of experiments conducted across multiple languages, which produced consistent insights, and by the fact that our analysis is not language-specific.
Furthermore, our in-depth analysis revealed no dependency on the programming language, as evidenced by the performance improvements observed when incorporating multiple languages into the training data.
Therefore, our findings are likely to generalize beyond Java and C/C++.

\section{Related Work}
\label{sec:related}

\subsection{Characteristics of Vulnerabilities and Security Patches}
A number of studies have examined the characteristics and effectiveness of security patches~\cite{li2017large,bandara2020fix,forootani2022exploratory}.
Li and Paxson~\cite{li2017large} analyzed over 4,000 bug fixes addressing more than 3,000 vulnerabilities. Utilizing the NVD to identify publicly disclosed vulnerabilities, their findings revealed several key insights. Compared to non-security bug patches, security patches tend to have a smaller impact on codebases, leading to more localized changes. They also observed that security issues often persist in codebases for years, with one-third introduced over three years before remediation. Additionally, 7\% of the security patches studied failed to fully resolve the targeted vulnerabilities.

Building on this work, Bandara et al.~\cite{bandara2020fix} analyzed 118,000 commits from 53 popular JavaScript projects hosted on GitHub. Their study revealed that in 82\% of the projects, security patches introduced new vulnerabilities in one-fifth of the cases. Notably, half of the total vulnerabilities in the studied projects originated from commits intended to fix previous vulnerabilities.

While these two studies explored the broader characteristics of security patches, Forootani et al.~\cite{forootani2022exploratory} focused on ``self-fixed'' vulnerabilities, where the developer who introduced the vulnerability also resolved it. 
Examining 1,752 security patches across projects written in C and PHP, covering over 17 types of CWEs, they found that most vulnerabilities were fixed by developers other than those who introduced them. 

Different from these previous studies, our work examines the limitations of using NVD-linked data as the sole training source for detecting security patches in real-world scenarios. 
Previous studies have primarily focused on disclosed security patches that can be linked to NVD entries. This approach may bias the results towards certain types of vulnerabilities that are more common in NVD entries, while overlooking unreported vulnerabilities that occur in real-world open-source projects, i.e., in-the-wild scenario.

By analyzing the differences between NVD-linked and in-the-wild patches, we highlight the importance of diverse datasets and realistic evaluations to enhance model robustness.

\subsection{Security Patch Detection Datasets}
In Section~\ref{sec:background}, we discuss two primary approaches for identifying security patches: those sourced from the NVD and those discovered in the wild. These two directions remain the mainstream methodologies for constructing security patch datasets.

SPI-DB~\cite{zhou2021spi} is a notable dataset that exemplifies this dual-source strategy. It comprises two components: (1) security patches crawled directly from the NVD and (2) manually verified security-related commits. However, SPI-DB focuses exclusively on four open-source software (OSS) projects, all of which are primarily written in C/C++.

In addition to identifying existing security patches with these two methods, an alternative approach for constructing security patch datasets is synthesizing security patches. 
PatchDB~\cite{wang2021patchdb} is a representative example of this method. PatchDB consists of three components: an NVD-based dataset, a \textit{wild-based} dataset, and a synthetic dataset. 
For the wild-based dataset, the authors developed a nearest-link search algorithm to identify GitHub commits that are most similar to the security patches in the NVD-based dataset. 
For the synthetic dataset, they programmatically modify critical statements at the source code level based on the two natural datasets, i.e., the NVD-based and wild-based datasets.
Similarly, PatchDB also only contains the security patches written in C/C++.

In contrast to the aforementioned datasets, the datasets adopted in our study are not limited to C/C++, specifically MoreFixes contains security patches written in 81 programming languages.
This broader coverage and analysis provide new insights into constructing diverse and realistic security patch datasets for training machine learning models.

\section{Conclusion and Future Work}
\label{sec:conclusion}

In this work, we provide empirical evidence that using security patches linked from the NVD as the single source of data is insufficient for training effective machine learning models. 
Our experiments show a significant decrease in performance, i.e., of up to 90\% in F1-score, when the models are applied to detect unreported vulnerabilities, e.g., silent fixes.
We utilized a dataset of manually validated security patches found by collecting commit data from open source Java and C/C++ repositories.
A deeper analysis reveals a different distribution of vulnerability types, i.e., CWEs, between NVD reports and the commits from JavaVFC projects, which may contribute to the lower performance of the models in a realistic evaluation setting.

Beyond differences in CWE distribution, there were also differences in the perplexity of the commit messages and the composition of the commits, e.g., different proportions of merge commits. 
Consequently, relying solely on NVD-related data introduces a bias toward disclosed vulnerabilities. 
This bias reduces the effectiveness of models when applied to commits unrelated to disclosed security patches, limiting their generalizability in real-world scenarios.

Our study showed that poor model generalization can be mitigated by constructing a new dataset that integrates data from multiple sources. 
Additionally, leveraging available information from existing commits such as commit messages and commit diffs, instead of relying solely on code changes has the potential to further improve patch identification performance.
Looking ahead, we will also investigate other methods to learn better representations of commits that may be more robust. We also plan to explore methods of identifying the specific parts of a commit that address a vulnerability, which would enable a more granular approach to training security patch detectors.

\section*{Acknowledgement}
This research / project is supported by the National Research Foundation, Singapore and Ministry of Digital Development \& Information under its Smart Nation and Digital Government Translational R\&D Grant (Award No: TRANS2026-TGC01). Any opinions, findings and conclusions or recommendations expressed in this material are those of the author(s) and do not reflect the views of National Research Foundation, Singapore.

\bibliographystyle{ACM-Reference-Format}
\bibliography{sample-base}

\appendix

\end{document}